\documentclass{article}

\usepackage{microtype}
\usepackage{graphicx}
\usepackage{subcaption}
\usepackage{booktabs} 
\usepackage{multirow} 
\usepackage[table]{xcolor} 

\usepackage{hyperref}

\usepackage[preprint]{icml2026}

\usepackage{amsmath}
\usepackage{amssymb}
\usepackage{mathtools}
\usepackage{amsthm}

\usepackage[capitalize,noabbrev]{cleveref}

\theoremstyle{plain}

\theoremstyle{definition}

\theoremstyle{remark}

\usepackage[textsize=tiny]{todonotes}
\usepackage{pifont}
\usepackage{enumitem}
\newcommand{\xmark}{\ding{55}}

\icmltitlerunning{ProAgentBench: A Benchmark for Proactive Service Agents}

\begin{document}

\twocolumn[
  \icmltitle{ProAgentBench: Evaluating LLM Agents for Proactive Assistance with Real-World Data
  }



  \icmlsetsymbol{equal}{*}

  \begin{icmlauthorlist}
    \icmlauthor{Yuanbo Tang}{equal,sigs,freeu}
    \icmlauthor{Huaze Tang}{equal,sigs,freeu}
    \icmlauthor{Tingyu Cao}{sigs,freeu}
    \icmlauthor{Lam Nguyen}{sigs,freeu}
    \icmlauthor{Anping Zhang}{sigs}
    \icmlauthor{Xinwen Cao}{sigs,freeu}
    \icmlauthor{Chunkang Liu}{sigs,freeu}
    \icmlauthor{Wenbo Ding}{sigs}
    \icmlauthor{Yang Li}{sigs}
  \end{icmlauthorlist}

  \icmlaffiliation{sigs}{Tsinghua Shenzhen Graduate School, Tsinghua University, Shenzhen, China}
  \icmlaffiliation{freeu}{FreeU Group (Open Collaborative AI Research Collective)}

  \icmlcorrespondingauthor{Wenbo Ding}{ding.wenbo@sz.tsinghua.edu.cn}
  \icmlcorrespondingauthor{Yang Li}{yangli.ai@ieee.org}

  \icmlkeywords{Proactive Agents, Long-term User Context, VLM Annotation, Privacy Protection, Human-Computer Interaction}

  \vskip 0.3in
]



\printAffiliationsAndNotice{}  

\begin{abstract}
Proactive agents that anticipate user intentions without explicit prompts represent a significant evolution in human-AI interaction, promising to reduce cognitive load and streamline workflows. 
However, existing datasets suffer from two critical deficiencies: (1) reliance on LLM-synthesized data that fails to capture authentic human decision-making patterns, and (2) focus on isolated tasks rather than continuous workflows, missing the {pre-assistance behavioral context} essential for learning proactive intervention signals.
To address these gaps, we introduce \textbf{ProAgentBench}, a rigorous benchmark for proactive agents in working scenarios. 
Our contributions include: (1) a hierarchical task framework that decomposes proactive assistance into timing prediction and assist content generation; (2) a privacy-compliant dataset with 28,000+ events from 500+ hours of real user sessions, preserving bursty interaction patterns (burstiness $B$=0.787) absent in synthetic data; and (3) extensive experiments that evaluates LLM- and VLM-based baselines. Numerically, we showed that long-term memory and historical context significantly enhance prediction accuracy, while real-world training data substantially outperforms synthetic alternatives.
We release our dataset and code at \url{https://anonymous.4open.science/r/ProAgentBench-6BC0}.
\end{abstract}

\begin{figure}[t]
  \centering
  \includegraphics[width=0.9\columnwidth]{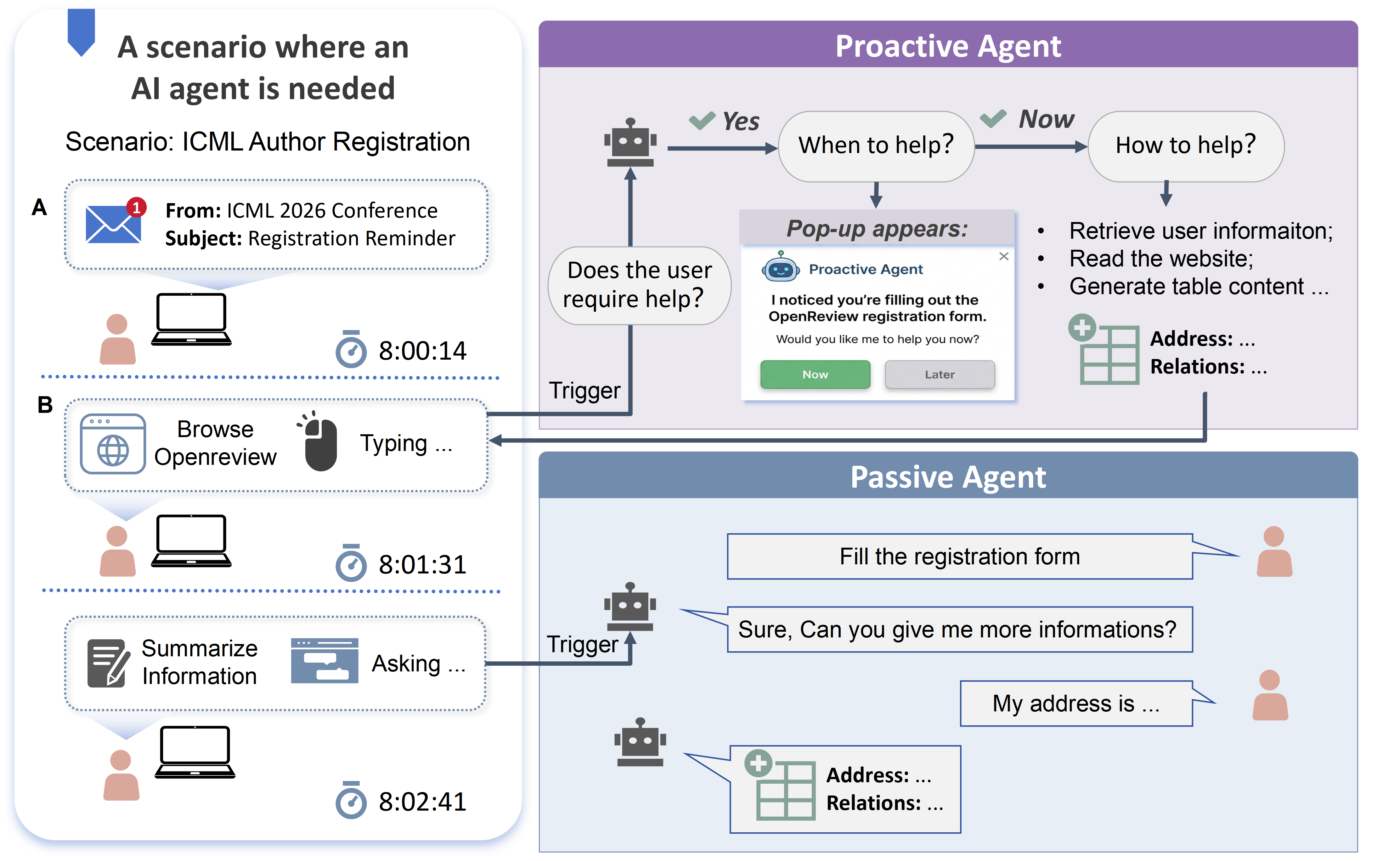}
  \caption{\textbf{Illustration of Proactive Agent Workflow.} The agent continuously monitors user screen activities and contextual signals. When assistance is needed, it proactively determines \textit{when} to intervene and \textit{how} to assist based on historical observations and user behavior patterns.}
  \label{fig:main_figure}
  \vspace{-20pt}
\end{figure}
\vspace{-10pt}
\vspace{-10pt}
\section{Introduction}

Recent breakthroughs in large language models (LLMs) and embodied agent research have shifted the paradigm of human-AI interaction from reactive instruction following to proactive assistance 
\citep{lu2024proactive,sun2025ppp,yang2025contextagent}. 
As is illustrated in Figure~\ref{fig:main_figure}, a proactive agent is defined as an AI system capable of perceiving environmental context, inferring user intentions without explicit prompts, and autonomously suggesting actions accordingly.
Unlike reactive agents that rely on explicit user commands, proactive agents aim to bridge the gap between observable behavior and latent user needs. 
Prior HCI research has established that poorly timed interruptions impose significant cognitive costs on productivity~\citep{mark2008cost}, requiring users to expend additional mental resources for task resumption and context recovery~\citep{iqbal2007disruption}. 
By anticipating user needs and providing timely, contextually appropriate assistance, proactive agents promise to reduce cognitive load and streamline task execution, enabling transformative productivity improvements in complex domains where human-AI collaboration is paramount.

Advancing proactive agents requires large-scale, high-quality datasets that capture authentic human-AI interaction patterns. 
However, existing datasets suffer from two critical deficiencies:
(1) \textbf{Lack of real-world data}: Current datasets predominantly rely on LLM-synthesized interactions, which fail to capture the stochastic nature of human decision-making and the bursty temporal patterns inherent in real workflows~\citep{mark2008cost}. Agents trained on such data exhibit brittleness when facing real-world ambiguity, while scalable collection of authentic data is impeded by privacy concerns.
(2) \textbf{Insufficient long-term data coverage}: Existing collections focus on isolated, short-duration tasks rather than continuous workflows, missing the \textit{pre-assistance behavioral context}, the critical signals of what users were doing {before} seeking help. This context is essential for learning timing and content to proactively intervene.

To address these gaps, we present \textbf{ProAgentBench}, the first rigorous benchmark designed to evaluate proactive agents in working scenarios. 
To tackle the \textbf{lack of real-world data}, we develop a privacy-compliant data collection pipeline that combines rule-based anonymization with human-in-the-loop review, enabling the safe collection of authentic user interactions at scale. 
Our dataset captures over 28,000 events from 500+ hours of continuous working sessions, preserving the bursty temporal patterns (burstiness $B=0.787$) that synthetic data fundamentally lacks.
To address \textbf{insufficient long-term coverage}, we record complete user work sessions rather than isolated tasks, explicitly capturing the \textit{pre-assistance behavioral context}, namely, what users were doing in the minutes before seeking AI help, that is critical for learning proactive intervention signals.
We then formalize a ``\textbf{When + How}'' hierarchical task framework that decomposes proactive assistance into two scientifically tractable problems: \textit{When to Assist} (binary classification of optimal intervention timing) and \textit{How to Assist} (generation of contextually appropriate content). 
This formalization enables systematic evaluation where each metric reflects real-world productivity impact: \textit{precision} quantifies interruption cost (low precision causes alert fatigue~\citep{alertfatigue2023}), while \textit{recall} measures need coverage (low recall fragments workflows~\citep{adamczyk2004interruptions}).
In summary, our work makes three key contributions that directly address the identified gaps:
\begin{itemize}
  \item We introduce \textbf{ProAgentBench}, the first rigorous benchmark for proactive agents providing a reusable paradigm for this emerging research area.
  \item We collect real-world human-AI interaction data with extensive user workflow logs, preserving authentic bursty interaction patterns and {pre-assistance behavioral context}, which are the critical signals preceding user needs.
  \item We conduct extensive experiments across diverse models and methods, revealing that both context and long-term memory significantly enhance prediction accuracy and real-world training data substantially outperforms synthetic data.
\end{itemize}

\section{Related Work}

\paragraph{Proactive Service Agents.} 
Recent advances in LLMs have catalyzed significant progress in proactive agent research. 
\citet{lu2024proactive} pioneered data-driven proactive agent training with ProactiveBench, collecting 6,790 real-world events and achieving 66.47\% F1-Score through reward modeling and fine-tuning. 
\citet{yang2025contextagent} introduced ContextAgent, which leverages multi-dimensional sensory perceptions from wearable devices to provide context-aware proactive assistance across 1,000 samples in daily life scenarios. 
In the mobile domain, \citet{yang2025fingertip} contributed FingerTip 20K, focusing on proactive task suggestions and personalized execution through long-term Android device interaction data. 
\citet{liu2025proactiveeval} proposed ProactiveEval, a unified evaluation framework that decomposes proactive dialogue into target planning and dialogue guidance across 328 environments. 
However, these works remain limited in data scale, scenario coverage, and privacy protection mechanisms. 
\vspace{-10pt}
\paragraph{Screen Recording Datasets for Computer Use.} 
The emergence of LLMs has driven demand for large-scale datasets capturing computer screen interactions. 
Pioneering efforts include Rico~\citep{deka2017rico}, which provided 72,000 mobile UI screenshots establishing foundations for data-driven interface analysis. 
For web environments, \citet{deng2023mind2web} introduced Mind2Web with over 2,000 tasks across 137 websites, while \citet{zhou2023webarena} released WebArena with fully functional environments, revealing that even state-of-the-art models achieve only modest success rates. 
Desktop coverage expanded through AssistGUI~\citep{gao2023assistgui} featuring professional software tasks, and OmniACT~\citep{kapoor2024omniact} with diverse desktop applications. 
\citet{chen2024gui} contributed GUI-World with 12,379 video recordings highlighting temporal information importance, while \citet{rawles2024androidworld} provided AndroidWorld with 116 parameterized mobile tasks. 

However, these datasets focus on task execution rather than proactive assistance, containing action sequences for predefined goals rather than organic user work patterns. 
Critically, they lack the \textit{pre-interaction context} that captures what users were doing \textit{before} seeking AI assistance, making it impossible to learn antecedent signals of user needs. 
They also lack the temporal density and privacy-preserving methodologies necessary for real PC work scenarios. 
Our ProAgentBench addresses this gap by capturing continuous workflows with both pre-LLM behavioral context and subsequent interaction events.

\begin{figure*}[t]
  \centering
  \begin{subfigure}[b]{0.32\textwidth}
    \centering
    \includegraphics[width=\linewidth]{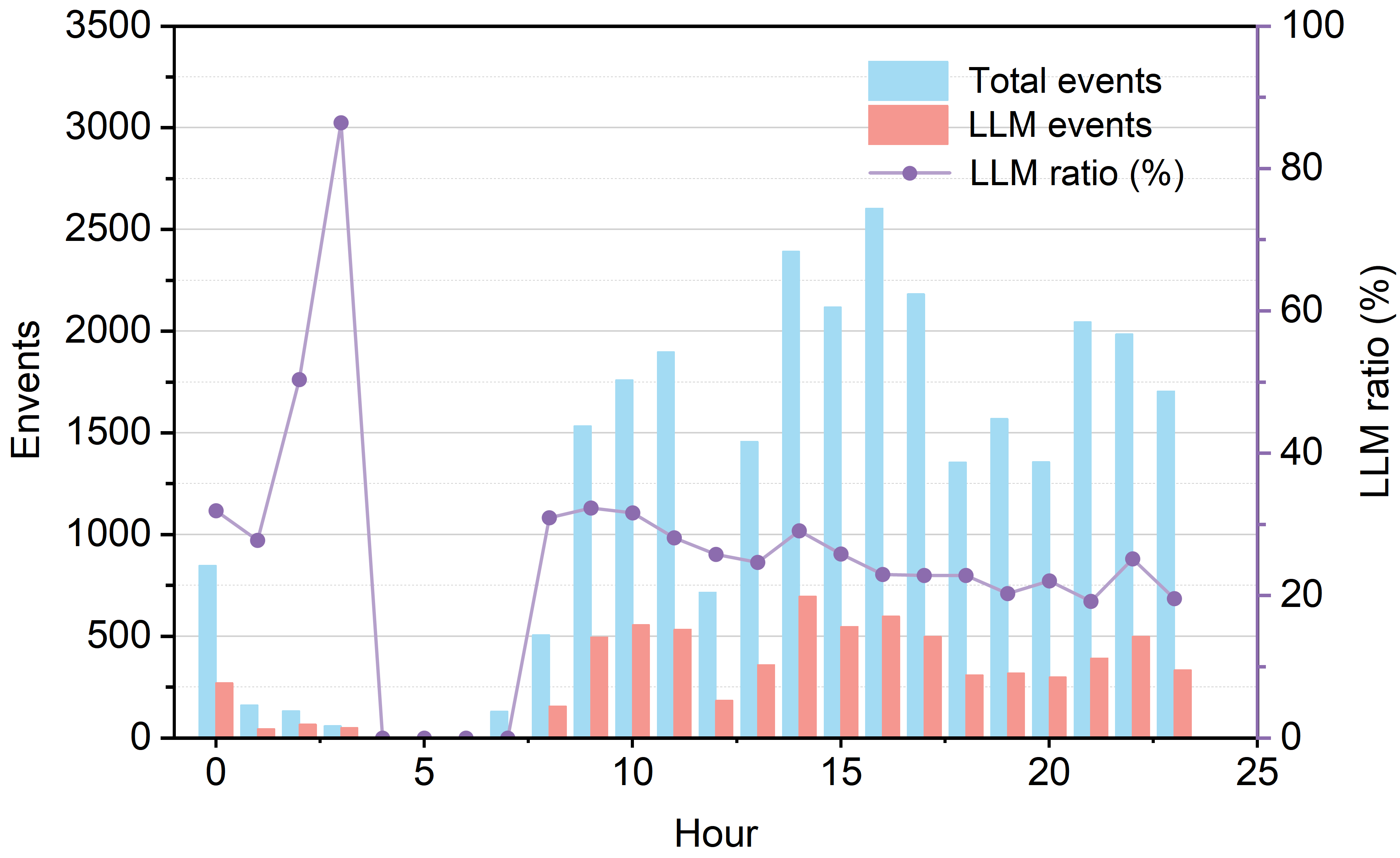}
    \caption{Weekday distribution.}
    \label{fig:weekday_total_llm_ratio}
  \end{subfigure}
  \hfill
  \begin{subfigure}[b]{0.32\textwidth}
    \centering
    \includegraphics[width=\linewidth]{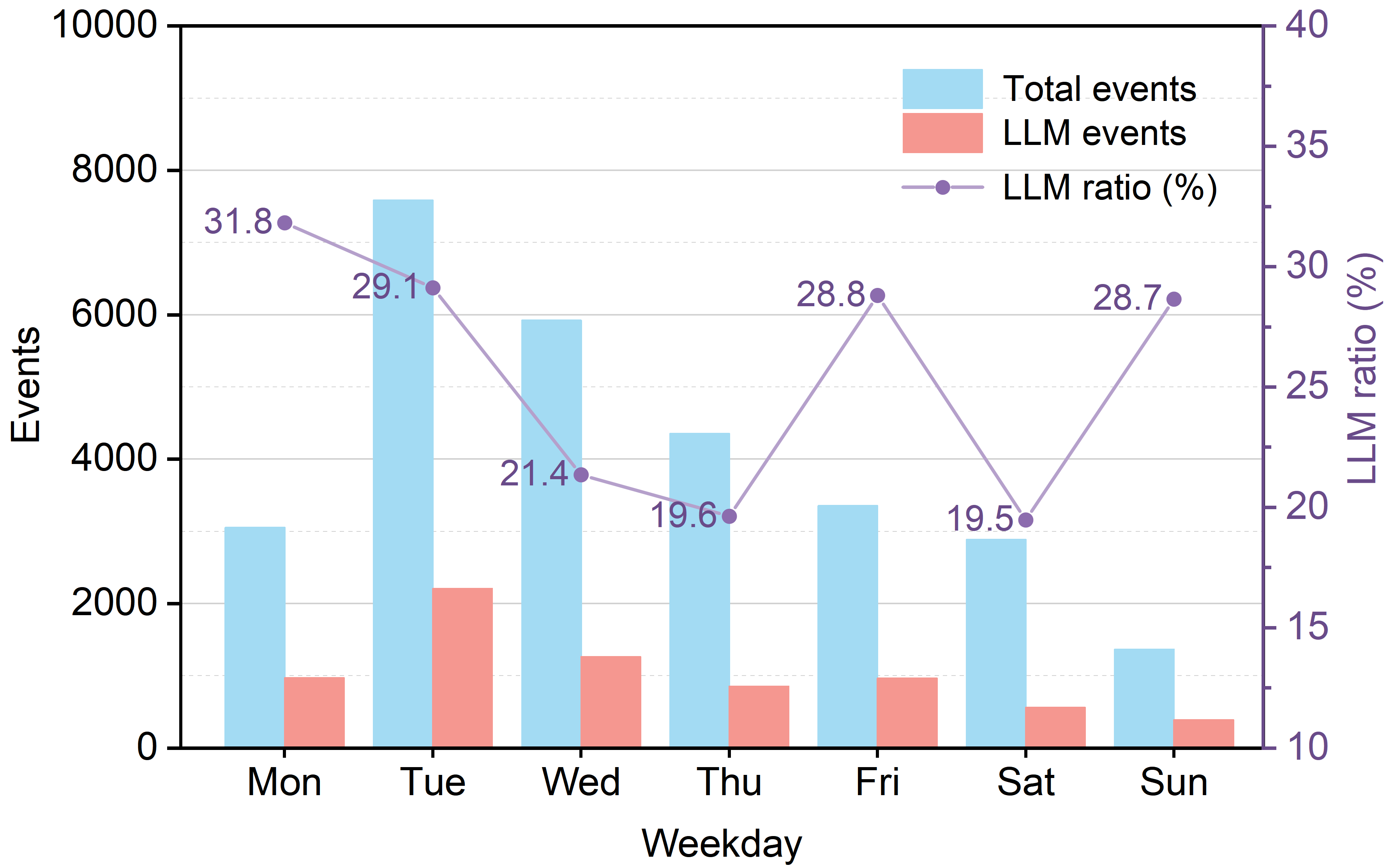}
    \caption{Hourly (0--23) distribution.}
    \label{fig:hour_total_llm_ratio}
  \end{subfigure}
  \hfill
  \begin{subfigure}[b]{0.28\textwidth}
    \centering
    \includegraphics[width=\linewidth]{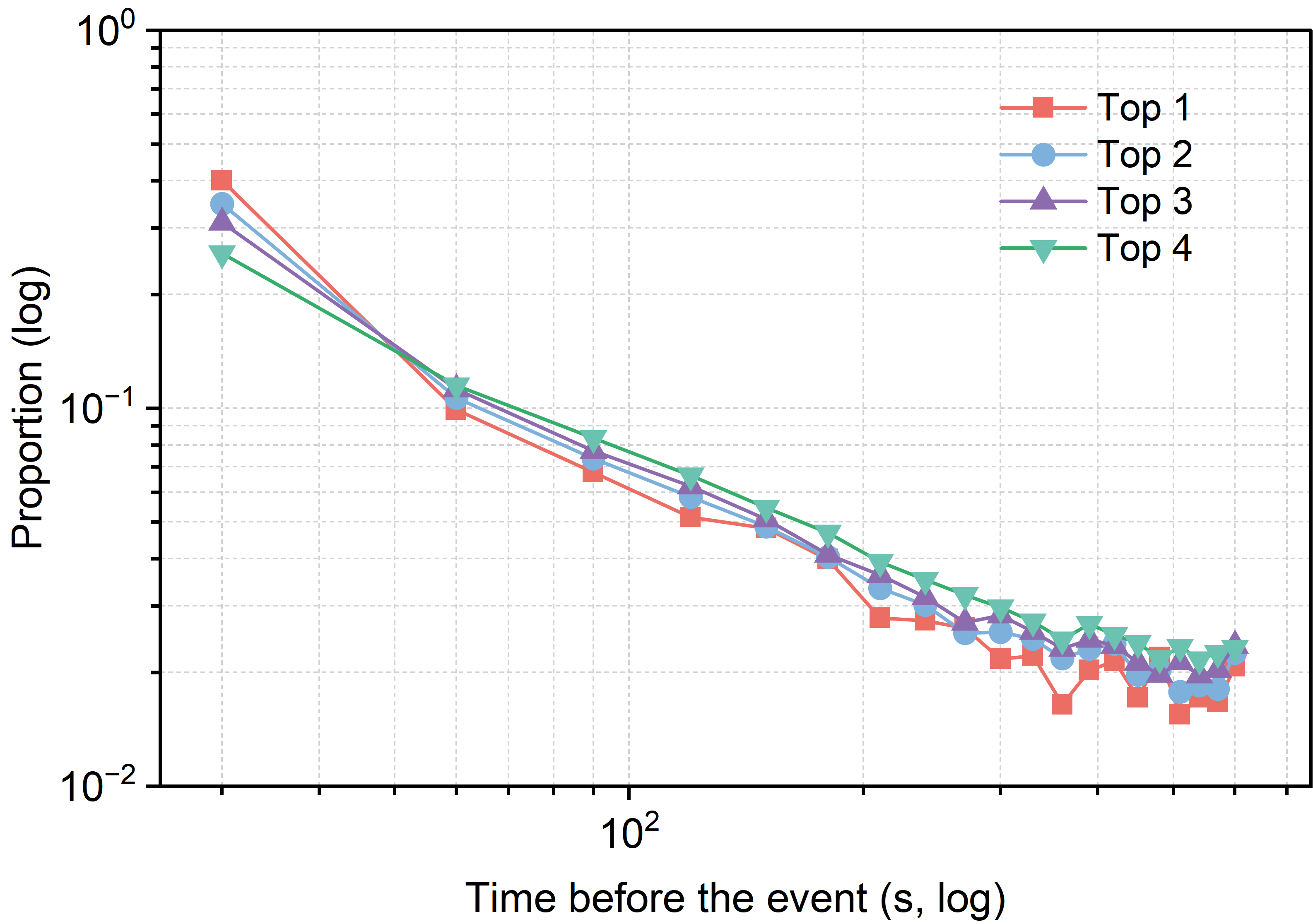}
    \caption{Top-$k$ similar screenshots.}
    \label{fig:topk_similar_screenshots}
  \end{subfigure}
  \caption{\textbf{Temporal distributions and context relevance.} We report total events, LLM events, and the LLM ratio across (a) weekdays and (b) hours of day, and (c) distribution of time-to-event for Top-1/3/5/10 nearest screenshots (log-log). Similarity computed using \texttt{qwen2.5-vl-embedding}.}
  \label{fig:llm_time_weekday_hour_topk}
  \vspace{-10pt}
\end{figure*}

\begin{figure}[t]
  \centering
  \begin{subfigure}[b]{0.48\columnwidth}
    \centering
    \includegraphics[width=\linewidth]{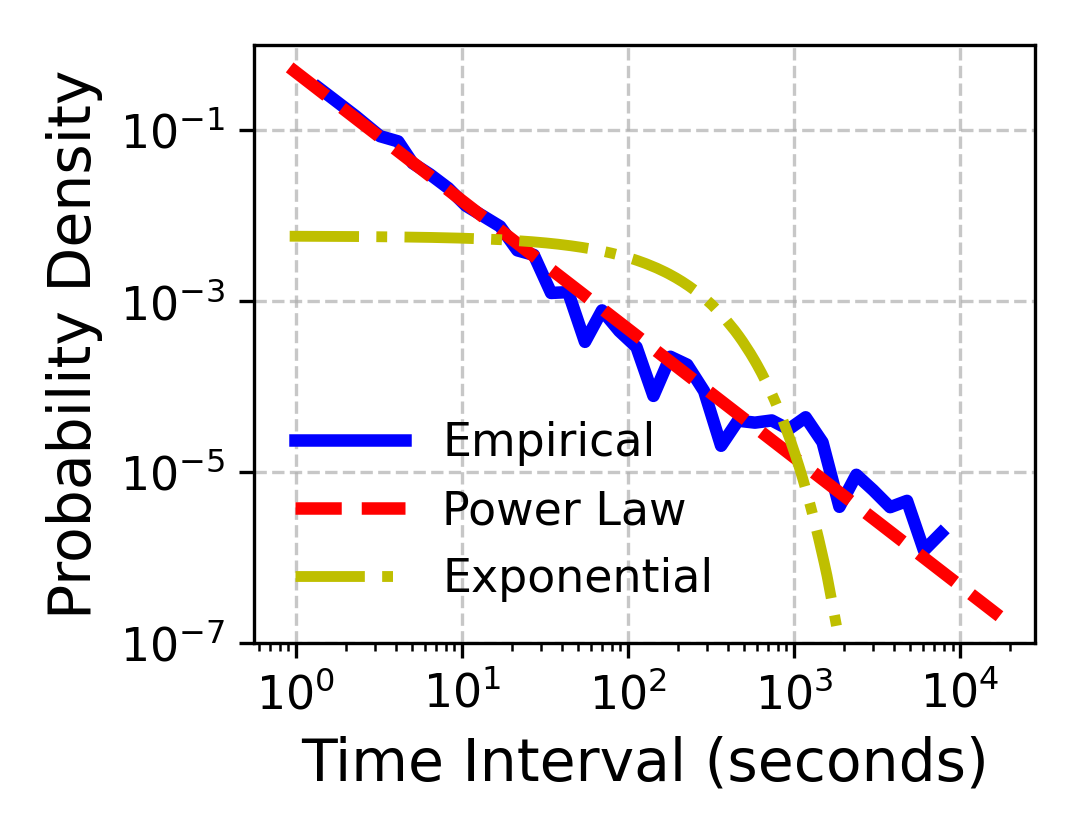}
    \caption{Human Data}
    \label{fig:human-data-statistics}
  \end{subfigure}
  \hfill
  \begin{subfigure}[b]{0.48\columnwidth}
    \centering
    \includegraphics[width=\linewidth]{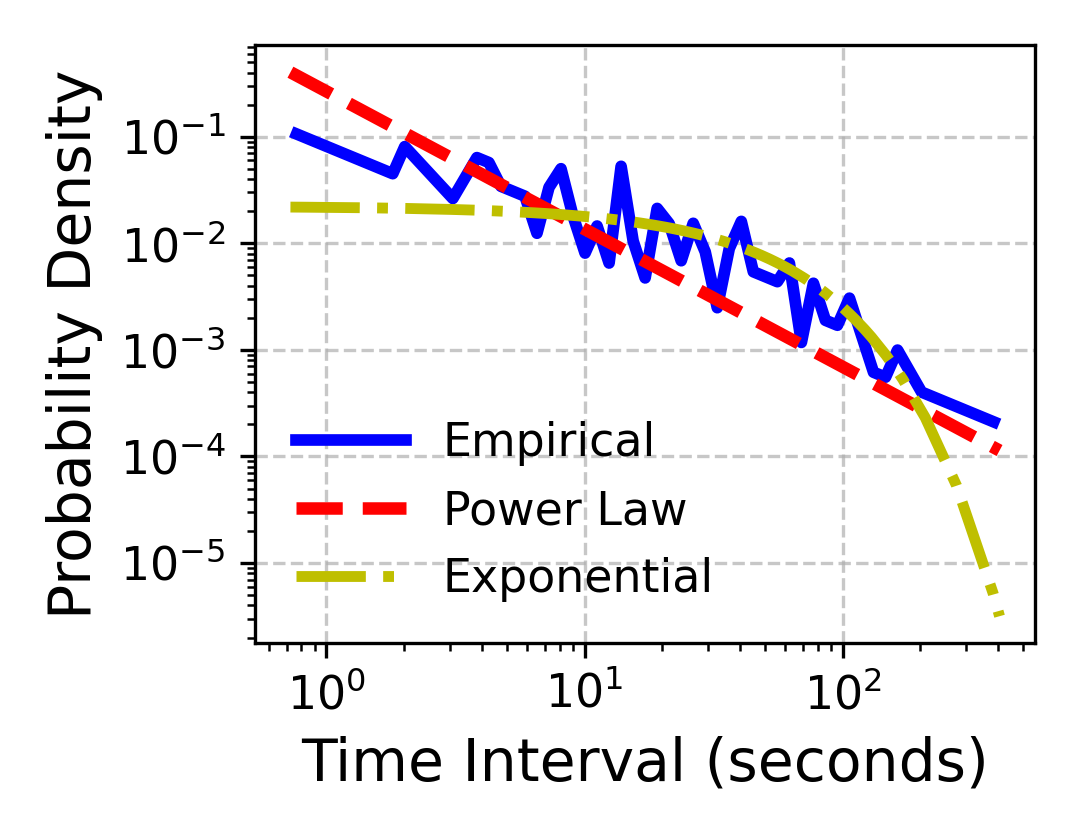}
    \caption{LLM Synthesized Data}
    \label{fig:llm-synthesize-data}
  \end{subfigure}
  \caption{Statistics of Human and LLM Synthesized Data}
  \label{fig:data-statistics}
  \vspace{-10pt}
\end{figure}

\section{Problem Definition and Formulation}
\label{sec:problem_definition}

In this work, we define the proactive agent as an intelligent system that continuously monitors the user's current screen snapshots in real-time and proactively initiates contact upon detecting a need for service. Specifically, the proactive agent maintains a rich context of the user's historical information. Through this context, the agent achieves two goals: (1) modeling the user's inherent behavioral patterns, and (2) deriving the contextual background of the user's current snapshots. Based on these foundations, the proactive agent determines whether assistance is required and infers the user's intent, subsequently providing concrete, intent-aligned assistance and services. In this section, we first specify the inputs of proactive agent. After that, we provide a formal definition of the Proactive Agent and the data structures involved in our system. We then outline the hierarchical pipeline that decomposes the proactive assistance problem into two sub-tasks: timing prediction (\textit{When to Assist}) and content generation (\textit{How to Assist}).
\vspace{-10pt}
\paragraph{Temporal Snapshot Sequence Inputs}
We define the input to a proactive assistance system as a temporal sequence of snapshots capturing user activities. At each time step $i$, the system captures a \textbf{snapshot} $S_i$ consisting of multiple raw modalities: 
(1) \textbf{Screen Image} $I_i$: A screenshot capturing the current visual state of the user's display, including application windows, UI elements, and on-screen content.
(2) \textbf{Timestamp} $\tau_i$: The precise time at which the snapshot was captured, enabling temporal analysis of user behavior patterns.
(3) \textbf{Application Metadata} $M_i$: Contextual information including the active application name, window title, and application category.
The historical observation sequence up to time $t$ is thus defined as $O_{1:t} = \{S_1, S_2, \ldots, S_t\}$, where each $S_i = (I_i, \tau_i, M_i)$. In addition to real-time observations, the system has access to user meta-information $U$ (e.g., role, preferences, and long-term memory derived from historical interactions).
Specifically, the user meta-information $U$ comprises:
(1) \textbf{Historical Information}: Records of the user's past interactions and long-term behavioral patterns, serving as a reference for contextual understanding.
(2) \textbf{User Profile}: A structured model of the user, including attributes such as occupation, domain expertise, and specific preferences, which guides personalized assistance.
\vspace{-10pt}
\paragraph{A Hierarchical Pipeline for Proactive Assistance}
Given the input sequence $O_{1:t}$ and user meta-information $U$, we design a two-stage pipeline that mirrors the natural decision process of an intelligent assistant. In the first stage (\textbf{When to Assist}), the agent continuously monitors user activities and determines the optimal moment to intervene, avoiding both premature interruptions that cause workflow disruption~\citep{mark2008cost} and missed opportunities that force users to manually seek assistance. Only when the first stage predicts a positive trigger does the second stage (\textbf{How to Assist}) activate, generating contextually appropriate assistance content. This hierarchical design reflects the real-world constraint that unnecessary assistance queries (false positives in Stage 1) incur user interruption costs, while missed needs (false negatives) result in degraded user experience.
\vspace{-10pt}
\paragraph{Task 1: When to Assist}

The interaction timing prediction is modelled as a {binary classification problem} that predicts whether proactive assistance is needed currently. Denoting the model as $f_{\text{when}}$, the prediction $B_t$ is given by
\begin{equation*}
B_t = f_{\text{when}}(U, O_{1:t}) \in \{0, 1\}
\end{equation*}
where $B_t = 1$ indicates that assistance is needed, and $B_t = 0$ indicates no intervention. The model $f_{\text{when}}$ is implemented with LLMs.
Our evaluation metrics are designed to directly reflect real-world productivity impact:
(1) \textbf{Accuracy}: Measures overall system reliability, directly correlating with user trust and long-term adoption willingness.
(2) \textbf{Precision}: Quantifies the rate of correct triggers among all interventions. Low precision leads to \textit{alert fatigue}, excessive false alarms causing users to ignore or disable assistance features, ultimately degrading productivity~\citep{alertfatigue2023}.
(3) \textbf{Recall}: Captures coverage of actual user needs. Low recall means missed assistance opportunities, forcing users to manually seek help and fragmenting their workflow~\citep{iqbal2007disruption}.
(4) \textbf{F1 Score}: Balances the trade-off between unnecessary interruptions (low precision) and missed opportunities (low recall), serving as a holistic measure of proactive system effectiveness.
\vspace{-10pt}
\paragraph{Task 2: How to Assist}

When $B_t = 1$, the agent generates assistance content $C_t$:
\begin{equation*}
C_t = f_{\text{how}}(U, O_{1:t}) \in \mathcal{V}
\end{equation*}
where $\mathcal{V}$ represents the natural language text space. The model $f_{\text{how}}$ is implemented with LLMs.
We evaluate the quality of generated assistance content using:
(1) \textbf{Intention Accuracy}: Classification accuracy for coarse intention categories (see Appendix \ref{app:intention_categories}). This metric reflects whether the agent correctly identifies the \textit{type} of assistance needed (e.g., information retrieval vs. code generation), which determines the relevance of the response.
(2) \textbf{Semantic Similarity}: Cosine similarity between predicted and real user query embeddings. This measures how well the generated content aligns with the user's actual query, directly impacting whether the assistance reduces or increases user effort.

\begin{table*}[t]
  \centering
  \caption{Comparison with representative proactive-agent datasets/benchmarks. ``Mixed'' indicates that the resource is constructed by combining real-world signals with synthetic/simulated components. ``LLM Queries'' refers to timestamped records of user interactions with AI assistants, providing direct evidence of when and how users seek assistance.}
  \label{tab:dataset-comparison}
  \small
  \setlength{\tabcolsep}{5pt}
  \resizebox*{0.9\textwidth}{!}{
  \begin{tabular}{lcccccc}
    \toprule
    \textbf{Dataset} & \textbf{Real-World Data} & \textbf{Pre-Event Logs} & \textbf{Long-Term Context} & \textbf{LLM Queries} & \textbf{Original Info} & \textbf{\#Events} \\
    \midrule
    \citet{lu2024proactive} & Mixed & \checkmark & \xmark & \checkmark & \xmark & 6,790 \\
    \citet{sun2025ppp} & Mixed & \xmark & \xmark & \checkmark & \xmark & 6,563 \\
    \citet{yang2025contextagent} & Mixed & \checkmark & \xmark & \checkmark & \checkmark & 1,000 \\
    \citet{yang2025fingertip} & \checkmark & \checkmark & \checkmark & \xmark & \checkmark & 21,437 \\
    \citet{liu2025proactiveeval} & Mixed & \checkmark & \checkmark & \checkmark & \xmark & 328 \\
    \midrule
    \textbf{ProAgentBench} (Ours) & \textbf{\checkmark} & \textbf{\checkmark} & \textbf{\checkmark} & \textbf{\checkmark } & \textbf{\checkmark} & \textbf{28,528} \\
    \bottomrule
  \end{tabular}}
\end{table*}

\section{Dataset Overview}

\subsection{Dataset Structure}
To answer the two challenges: Impact of real-world data and long-term user context, we build up a dataset from real users with long-term user logs.
We compare it with existing proactive assistance benchmarks in Table~\ref{tab:dataset-comparison}. Most of the existing benchmarks rely on synthetic training data or simulated environments and lack authentic long-term user context. For instance, \citet{lu2024proactive} uses synthetic training scenarios, \citet{sun2025ppp} employs simulated user feedback, and \citet{yang2025contextagent} relies on fabricated scenarios. Such reliance limits their ability to capture the natural temporal dynamics of real-world workflows. In contrast, our dataset is derived entirely from continuous, real-world user activity logs, providing both pre-assistance behavioral traces and long-term context. This authentic, large-scale data enables robust evaluation of both \textit{when} and \textit{how} to assist within a unified, realistic framework.

\begin{figure*}[ht]
  \centering
  \includegraphics[width=0.75\textwidth]{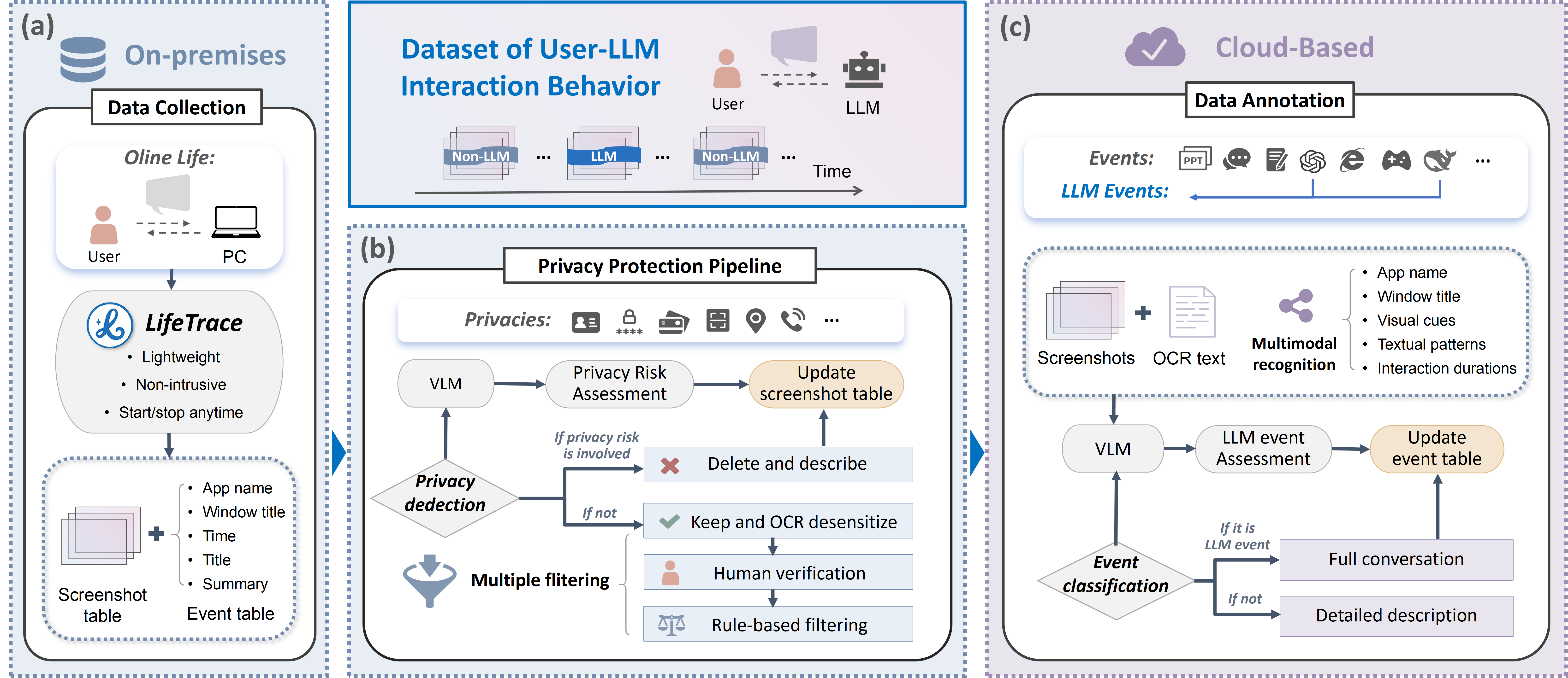}
  \caption{Data Collection Pipeline Overview. The figure illustrates the end-to-end data collection process, including screenshot capture, metadata synchronization, privacy filtering, and storage workflow.}
  \label{fig:data-collection-pipeline}
\end{figure*}

\subsection{Dataset Statistics and Analysis}
\label{sec:dataset_statistics}

\paragraph{User Profile Statistics} Our dataset primarily comprises student participants, spanning late undergraduate years and master's programs, and covers diverse academic backgrounds (e.g., computer science, electronic information, finance, biomedicine, energy, and translation). We collect 28,528 total events, among which 7,222 are LLM-related (\(\approx 25.3\%\)).
To characterize usage purposes, we categorize LLM interactions by event semantics, including Information Retrieval (35.10\%), Knowledge Q\&A (20.42\%), Data Analysis (9.17\%), Code Programming (8.72\%), and Content Generation (6.94\%).
From an application perspective, LLM-related events occur predominantly in web browsers (62.53\%), and appear consistently in file management tools (11.34\%), IDEs (9.25\%), and office software (9.14\%). Finally, at the platform level, identifiable providers are led by DeepSeek (23.62\%)~\citep{deepseekv3} and Gemini (18.69\%)~\citep{team2024gemini}, alongside ChatGPT, Cursor, Doubao, and Kimi~\citep{team2025kimi}.
\vspace{-10pt}
\paragraph{Temporal Usage Statistics} To further characterize temporal usage patterns, we summarize weekday-level and hour-of-day distributions of total events, LLM events, and the LLM ratio; see \Cref{fig:weekday_total_llm_ratio,fig:hour_total_llm_ratio}. We also analyze pre-event context by retrieving the most semantically similar screenshots within a 10-minute window before each LLM event, as illustrated in \Cref{fig:topk_similar_screenshots}. Specifically, we compute similarity between the LLM conversation-text embedding and screenshot image embeddings using Qwen2.5-VL-embedding~\citep{wang2024qwen2vl}, where top-$k$ denotes the set of the $k$ most similar screenshots. We observe the existence of a power-law relationship in the temporal distribution of relevant context. This finding underscores the critical importance of incorporating historical data, as key triggers for user needs are often buried in earlier interactions rather than being immediately adjacent to the current event.
\vspace{-10pt}
\paragraph{Bursty Human-LLM Interaction.}
Let $\{t_i\}_{i=1}^{N}$ denote the time stamps of observed interactions, sorted such that $t_{i+1}\ge t_i$.
We define the inter-event time (IET) as $\Delta t_i = t_{i+1}-t_i$, where $i=1,\ldots,N-1$.
To quantify whether the IETs exhibit a heavy tail, we fit a power law on the tail of $\{\Delta t_i\}$.
Specifically, we assume $p(\Delta t) \propto (\Delta t)^{-\alpha}$, and estimate the exponent $\alpha$ using maximum likelihood.
We compare the power law fit against an exponential alternative using a log-likelihood ratio test, where a positive log-likelihood ratio indicates that the power law provides a better fit, and a negative value favors the exponential model.
In addition, we report the burstiness score $B$ using IET~\citep{goh2008burstiness}:
\begin{equation}
    B = \frac{\sigma_{\Delta t}-\mu_{\Delta t}}{\sigma_{\Delta t}+\mu_{\Delta t}},
\end{equation}
where $\mu_{\Delta t}$ and $\sigma_{\Delta t}$ denote the sample mean and sample standard deviation of $\{\Delta t_i\}$, respectively.
By construction, $B\in[-1,1]$, with larger values indicating stronger temporal clustering and more bursty interaction patterns.
In the human interaction records, the IET distribution is heavy-tailed and is well fitted by a power law with exponent $\alpha=1.50$ (Fig.~\ref{fig:human-data-statistics}). A log-likelihood ratio test strongly supports the power law over the exponential model (log-likelihood ratio $=2951.48$, $p=7.83\times 10^{-100}$). The burstiness score is also high ($B=0.787$), consistent with many short gaps and a few long gaps.
For the synthetic data, we keep the same candidate time points as in the human records and let the LLM decide at each time point whether to interact. Under this setting, the IET distribution becomes closer to an exponential form on a log-log plot (Fig.~\ref{fig:llm-synthesize-data}). In this case, the likelihood ratio test indicates that the exponential model fits better (log-likelihood ratio $=-59.36$, $p=8.37\times 10^{-7}$), and the burstiness score drops to $B=0.166$. This suggests that even with realistic candidate time points, the LLM does not naturally reproduce the bursty timing in human behavior.

\begin{table*}[t]
  \centering
  \caption{Performance comparison of prompt-based methods across different models. We report metrics for both the \textit{When to Assist} task (Accuracy, Precision, Recall, F1 Score) and the \textit{How to Assist} task (Intention Accuracy, Semantic Similarity). The best result in each column is \textbf{bolded}, and the second best is \underline{underlined}.}
  \label{tab:prompt_based_results}
  \resizebox{0.8\textwidth}{!}{
  \begin{tabular}{llcccccc}
    \toprule
    \multirow{2}{*}{\textbf{Model}} & \multirow{2}{*}{\textbf{Method}} & \multicolumn{4}{c}{\textbf{When to Assist}} & \multicolumn{2}{c}{\textbf{How to Assist}} \\
    \cmidrule(lr){3-6} \cmidrule(lr){7-8}
    & & Accuracy & Precision & Recall & F1 Score & Intent. Acc. & Sem. Sim. \\
    \midrule
    \rowcolor{gray!15} \multicolumn{8}{l}{\textit{Closed-Source Models}} \\
    \multirow{3}{*}{GPT-4o-mini} 
      & Zero-shot & 54.9\% & 52.7\% & 96.2\% & 68.1\% & 28.4\% & 0.280 \\
      & CoT & 55.7\% & 55.6\% & \textbf{99.5\%} & \textbf{71.3\%} & 30.5\% & \underline{0.298} \\
      & Self-Consistency & 55.2\% & 52.8\% & 96.0\% & 68.2\% & 28.2\% & 0.280 \\
    \cmidrule(lr){1-8}
    \multirow{3}{*}{Qwen3-Max} 
      & Zero-shot & 59.3\% & 55.5\% & 93.4\% & 69.7\% & 36.3\% & 0.285 \\
      & CoT & 59.8\% & \underline{59.6\%} & 72.5\% & 65.4\% & \textbf{38.2\%} & \textbf{0.305} \\
      & Self-Consistency & 59.5\% & 55.7\% & 93.5\% & \underline{69.9\%} & 36.2\% & 0.285 \\
    \cmidrule(lr){1-8}
    \multirow{3}{*}{Deepseek-V3.2} 
      & Zero-shot & \textbf{64.4\%} & \underline{60.8\%} & 81.1\% & 69.5\% & 36.5\% & 0.276 \\
      & CoT & \underline{61.1\%} & \textbf{60.9\%} & 86.6\% & \textbf{71.3\%} & 35.0\% & 0.287 \\
      & Self-Consistency & \textbf{64.4\%} & \underline{60.8\%} & 81.1\% & 69.6\% & 36.5\% & 0.276 \\
    \cmidrule(lr){1-8}
    \multirow{3}{*}{Qwen3-VL-Plus} 
      & Zero-shot & 53.0\% & 51.6\% & 97.0\% & 67.4\% & \underline{37.1\%} & 0.286 \\
      & CoT & 53.5\% & 54.9\% & 61.3\% & 57.9\% & 34.4\% & \textbf{0.305} \\
      & Self-Consistency & 53.1\% & 51.7\% & 97.0\% & 67.4\% & 36.7\% & 0.286 \\
    \midrule
    \rowcolor{gray!15} \multicolumn{8}{l}{\textit{Open-Source Models}} \\
    \multirow{3}{*}{Llama-3.1-8B-Instruct} 
      & Zero-shot & 57.3\% & 54.7\% & 85.7\% & 66.7\% & 32.3\% & 0.275 \\
      & CoT & 50.8\% & 50.4\% & \underline{99.0\%} & 66.8\% & 29.1\% & 0.294 \\
      & Self-Consistency & 58.8\% & 56.7\% & 85.3\% & 68.1\% & 32.5\% & 0.274 \\
    \cmidrule(lr){1-8}
    \multirow{3}{*}{Qwen3-VL-8B-Instruct} 
      & Zero-shot & 51.7\% & 50.9\% & 94.4\% & 66.1\% & 35.3\% & 0.276 \\
      & CoT & 41.0\% & 32.7\% & 17.1\% & 22.4\% & 34.1\% & 0.277 \\
      & Self-Consistency & 52.9\% & 51.8\% & 93.6\% & 66.7\% & 35.7\% & 0.274 \\
    \bottomrule
  \end{tabular}
  }
\end{table*}
\section{Data Collection, Privacy Protection, and Automatic Annotation}
We employ LifeTrace\footnote{\url{https://github.com/FreeU-group/LifeTrace}} to collect real-world computer usage data. To construct a high-quality, privacy-compliant dataset, we design a pipeline consisting of three main phases: data collection, privacy protection, and automatic annotation, as is illustrated in Figure~\ref{fig:data-collection-pipeline}.

\subsection{Data Collection and Quality Control}
We collect user screen screenshots at 1Hz and synchronized application usage logs. Continuous user activities are automatically segmented into discrete events based on application switching. To ensure dataset quality, we implement a multi-layered filtering mechanism that excludes invalid events (e.g., extremely short duration or missing screenshots) and applies hash-based deduplication. Detailed setups and filtering criteria are provided in Appendix~\ref{app:data_collection}.

\subsection{Privacy Protection}
We prioritize user privacy through a rigorous three-stage process combining automated detection and human oversight. First, a VLM performs preliminary screening to identify sensitive visual and textual content. Second, we implement a human-in-the-loop mechanism where volunteers review and have final control over data retention. Finally, a rule-based filtering system acts as a safety net to catch remaining sensitive patterns. High-risk data is permanently deleted. The detailed privacy protocol is described in Appendix~\ref{app:privacy}.

\subsection{Automatic LLM Event Annotation}
To identify LLM interaction scenarios, we develop an event-level automatic annotation workflow. Unlike independent screenshot analysis, our approach aggregates multi-modal context (image sequences, OCR text, and metadata) within an event window. We utilize Qwen3-VL-Plus~\citep{qwen3} to classify LLM platforms, interaction types, and extract conversation history. Specific prompt designs and annotation logic are detailed in Appendix~\ref{app:annotation}.

\section{Experiments and Results}

\subsection{Experimental Setup}
We conduct experiments simulating realistic user workflows. We utilize all interaction events occurring within the past 5 minutes as the historical context for each prediction step. This window captures the immediate workflow continuity while minimizing noise from stale activities. We evaluate a diverse set of state-of-the-art Large Language Models (LLMs) and Vision-Language Models (VLMs), including both closed-source (GPT-4o-mini~\citep{openai2024gpt4o}, Qwen3-VL-Plus, Qwen3-Max~\citep{qwen3}, Deepseek-V3.2~\citep{deepseekv3}) and open-source (LLaMA3.1-8B-Instruct~\citep{llama3}, Qwen3-VL-8B-Instruct~\citep{qwen3}) variants. For all model inferences, we adhere to the default hyperparameters provided by the respective model APIs or official repositories (e.g., temperature, top-p) to ensure a fair and reproducible baseline comparison.

\subsection{Data Splits and Evaluation Protocol}
We implement a data splitting and evaluation protocol. 
First, we isolate each user's interaction history to prevent any cross-user information interference. 
Second, we employ {time-based splits} to mimic real-world deployment scenarios and avoid temporal data leakage.
Finally, for the interaction timing prediction task, we carefully select non-assistance moments that are contextually similar to actual assistance triggers. This strategy filters out trivial negatives (such as periods of inactivity), forcing the model to distinguish between subtle differences in user needs and providing a more realistic benchmark for proactive assistance.

\subsection{Base Results: Performance of Prompt-based Methods for Different Models}
We first evaluate the effectiveness of state-of-the-art LLMs on proactive assistance using three prompt-based baselines we designed for this task: \textbf{Zero-shot}, \textbf{Chain-of-Thought (CoT)}~\citep{wei2022chain}, and \textbf{Self-Consistency}~\citep{wang2023selfconsistency}. While CoT and Self-Consistency are general prompting strategies, we adapt them with task-specific prompt designs tailored to the proactive assistance setting (see Appendix~\ref{sec:vlm_prompts} for detailed prompt templates). Table~\ref{tab:prompt_based_results} presents the comprehensive results across both tasks.

\textbf{Zero-shot Performance.}
Among all evaluated models, Deepseek-V3.2 achieves the highest accuracy of 64.4\% on the \textit{When to Assist} task. Notably, closed-source models generally outperform their open-source counterparts. For the \textit{How to Assist} task, Qwen3-VL-Plus achieves the best intention prediction accuracy of 37.1\%. However, the semantic similarity scores remain relatively low across all models (ranging from 0.275 to 0.286), indicating that even state-of-the-art LLMs struggle to generate assistance content that closely matches user expectations.

\textbf{Impact of Chain-of-Thought Prompting.}
We observe that CoT prompting yields mixed results depending on model capacity. For larger models, CoT improves \textit{when to assist} performance. However, for smaller open-source models, CoT can be detrimental. This aligns with recent findings that CoT degrades performance on tasks involving implicit pattern recognition rather than explicit logical deduction \cite{liu2025mind, zheng2025curse}. Our analysis reveals that CoT amplifies models' inherent behavioral tendencies: in Deepseek-V3.2 and LLaMA3.1-8B, CoT shifts decision boundaries toward aggressive triggering (higher Recall), while in Qwen3-VL-8B, it induces excessive conservatism (lower Recall).
We further observe that CoT tends to overthink simple scenarios, imagining future problems rather than assessing what the user actually needs in the present,  as illustrated in Figure~\ref{fig:cot_failure}.
On the \textit{How to Assist} task, CoT provides modest improvements in semantic similarity, indicating that structured reasoning helps models better articulate assistance content.


\textbf{Self-Consistency Analysis.}
Self-Consistency sampling demonstrates stable but limited improvements over Zero-shot baselines. For instance, Llama-3.1-8B-Instruct improves from 57.3\% to 58.8\% accuracy, while Qwen3-VL-8B-Instruct improves from 51.7\% to 52.9\%. The F1 scores remain largely consistent across models. Notably, Self-Consistency does not significantly enhance intention prediction accuracy or semantic similarity, suggesting that the bottleneck lies in the models' fundamental understanding of user needs rather than output consistency.

\textbf{Key Observations.}
Our experiments reveal several important findings: (1) The proactive assistance task remains challenging, with the both best accuracy on timing prediction and intention prediction remains low; (2) The gap between timing prediction accuracy and intention prediction accuracy suggests that \textit{when to assist} is easier to determine than \textit{how to assist}; (3) Advanced prompting techniques may harm performance; (4) The low semantic similarity scores indicate substantial room for improvement in generating contextually appropriate assistance.
\vspace{-5pt}
\subsection{Research Question 1: Impact of Historical Observation Sequence Length}
To systematically evaluate the impact of historical information on proactive assistance, we conduct an ablation study on the {historical observation sequence length} $O_{1:t}$. This parameter controls the temporal range of user interaction logs provided to the model. We specifically investigate six distinct time settings: \textbf{10 seconds}, \textbf{30 seconds}, \textbf{1 minute}, \textbf{2 minutes}, \textbf{5 minutes}, and \textbf{10 minutes}. These settings allow us to analyze how the model's performance with different context ranging from immediate short-term history to more extended behavioral sequences.

\begin{figure}[t]
  \centering
  \begin{subfigure}[b]{0.49\columnwidth}
    \centering
    \includegraphics[width=\linewidth]{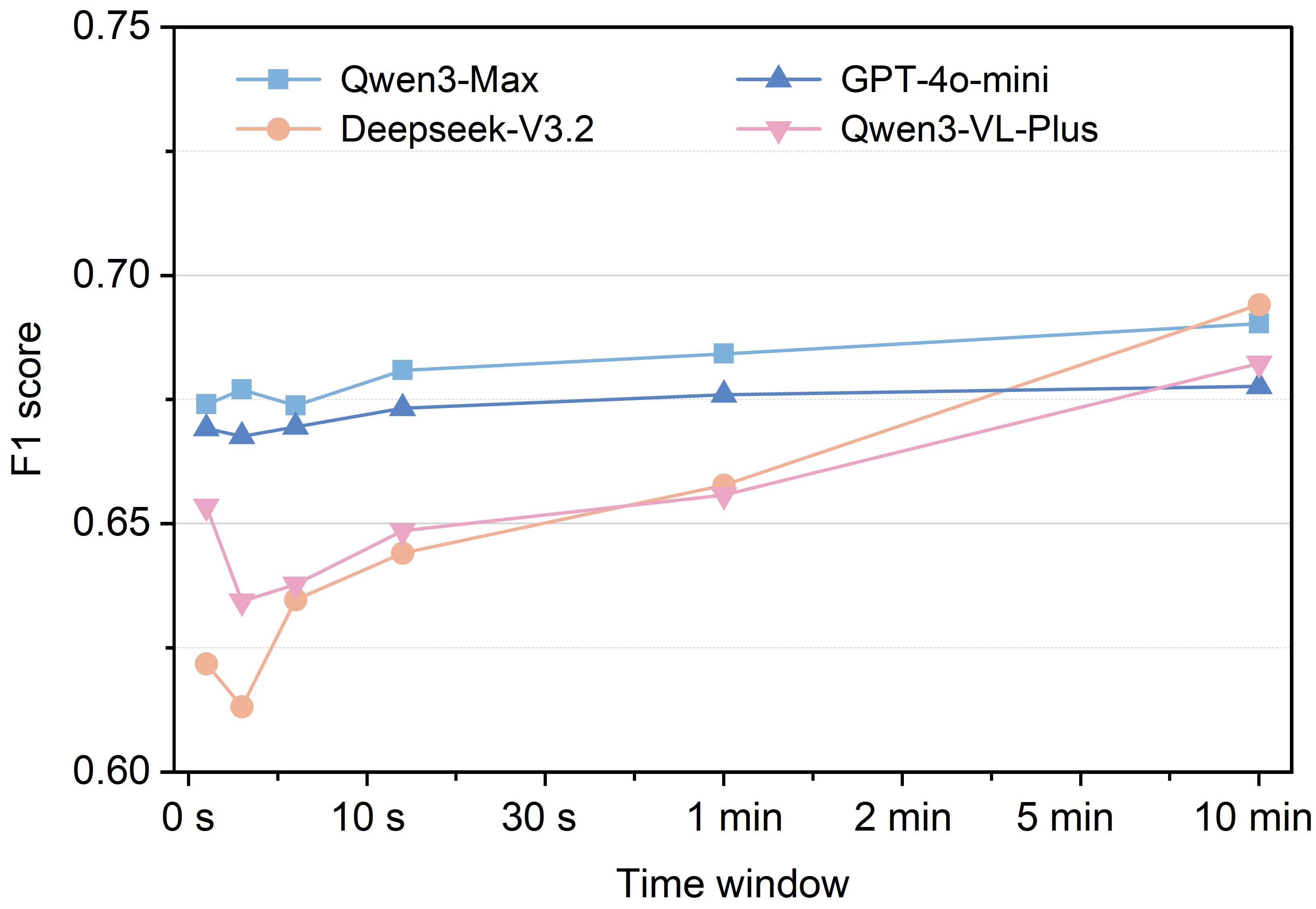}
    \caption{F1 score of Timing Task.}
    \label{fig:time_window_acc}
  \end{subfigure}
  \hfill
  \begin{subfigure}[b]{0.49\columnwidth}
    \centering
    \includegraphics[width=\linewidth]{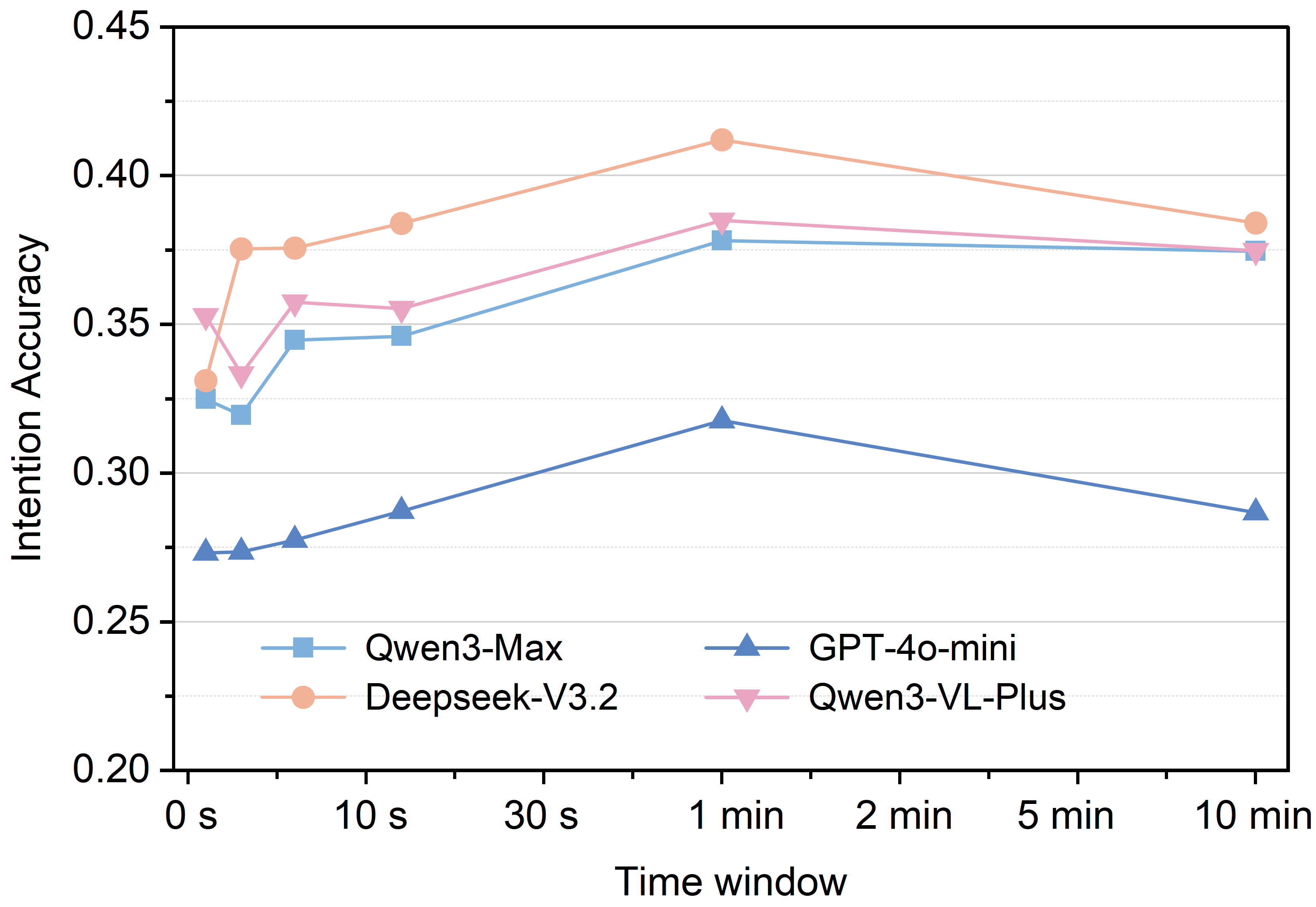}
    \caption{Intention Prediction Acc.}
    \label{fig:time_window_intention_acc}
  \end{subfigure}
  \caption{\textbf{Impact of Historical Context Length.} We evaluate the performance of proactive assistance across different time window sizes (from 30s to 10m). (a) F1 score on the ``When to Assist'' task. (b) Intention accuracy on the ``How to Assist'' task.}
  \label{fig:impact_context_length}
\end{figure}

We evaluate the performance using closed-source models. As illustrated in Figure~\ref{fig:impact_context_length}, we observe that both tasks benefit from longer historical context, though with different magnitudes. For the \textit{when to assist} task, extending the context window leads to gradual improvements in F1 score (Figure~\ref{fig:time_window_acc}), indicating that richer behavioral history helps the model better distinguish assistance-needed moments from normal activities. Similarly, for the \textit{how to assist} task, intention prediction accuracy also improves as the context window expands (Figure~\ref{fig:time_window_intention_acc}). Notably, intention accuracy exhibit diminishing returns beyond the 5-minute mark, with marginal gains observed between 5 and 10 minutes. This suggests that a 5-minute context window strikes an effective balance between capturing sufficient behavioral context and computational efficiency.

This finding aligns well with the semantic relevance analysis presented in Figure~\ref{fig:topk_similar_screenshots}. While the majority of highly relevant events appear within a short time window immediately preceding the LLM interaction, the relevance distribution exhibits a pronounced long-tail effect. Specifically, although the most semantically similar events cluster within the first few minutes, a non-negligible portion of contextually important information spans beyond this immediate horizon. This long-tail characteristic explains why longer context windows are particularly beneficial for the \textit{How to Assist} task: accurately inferring user intention often requires capturing sporadic but critical historical cues that may occur several minutes prior to the current interaction, even if they are temporally distant.

\vspace{-5pt}
\subsection{Research Question 2: Impact of Long-Term User Context}
\vspace{-5pt}

To evaluate the long-term user context in proactive assistance, we investigate the impact of incorporating long-term user behavior patterns. We introduce several \textbf{memory-based methods} that allow the agent to reference historical interaction data. Specifically, we implement and compare three distinct memory retrieval and organization strategies: (1) \textbf{Retrieval-Augmented Generation (RAG)}~\citep{Lewis2020}, which retrieves via semantic similarity; (2) \textbf{Knowledge Graph (KG)}~\citep{personax2025}, which structures user habits into a relational graph; and (3) \textbf{Clustering}, inspired by the PersonaX approach~\citep{Shi2025}, which categorizes user behaviors into distinct archetypes. 

We observe that: (1) Incorporating long-term user behavior patterns via memory-based methods significantly enhances the effectiveness of personalized AI in proactive assistance, with Knowledge Graph (KG) emerging as the most optimal strategy. Among the three memory retrieval and organization approaches, KG achieves the most substantial performance improvement over the Zero-shot baseline, increasing overall Accuracy by 11.8\% (from 0.537 to 0.601), Intention Accuracy by 26.9\% (from 0.312 to 0.396), and F1 Score by 6.1\% (from 0.675 to 0.716); (2) RAG demonstrates moderate effectiveness in leveraging historical interaction data, providing incremental improvements compared to the baseline without user behavior modeling. Specifically, RAG Memory-Based method improves Accuracy by 2.4\% (reaching 0.550), Intention Accuracy by 6.3\% (reaching 0.332), and maintains a stable F1 Score with a 0.8\% increase (reaching 0.681), indicating its ability to reference relevant historical snippets effectively but with limited reasoning capability compared to KG.

\begin{figure}[t]
  \centering
  \begin{subfigure}[b]{0.49\columnwidth}
    \centering
    \includegraphics[width=\linewidth]{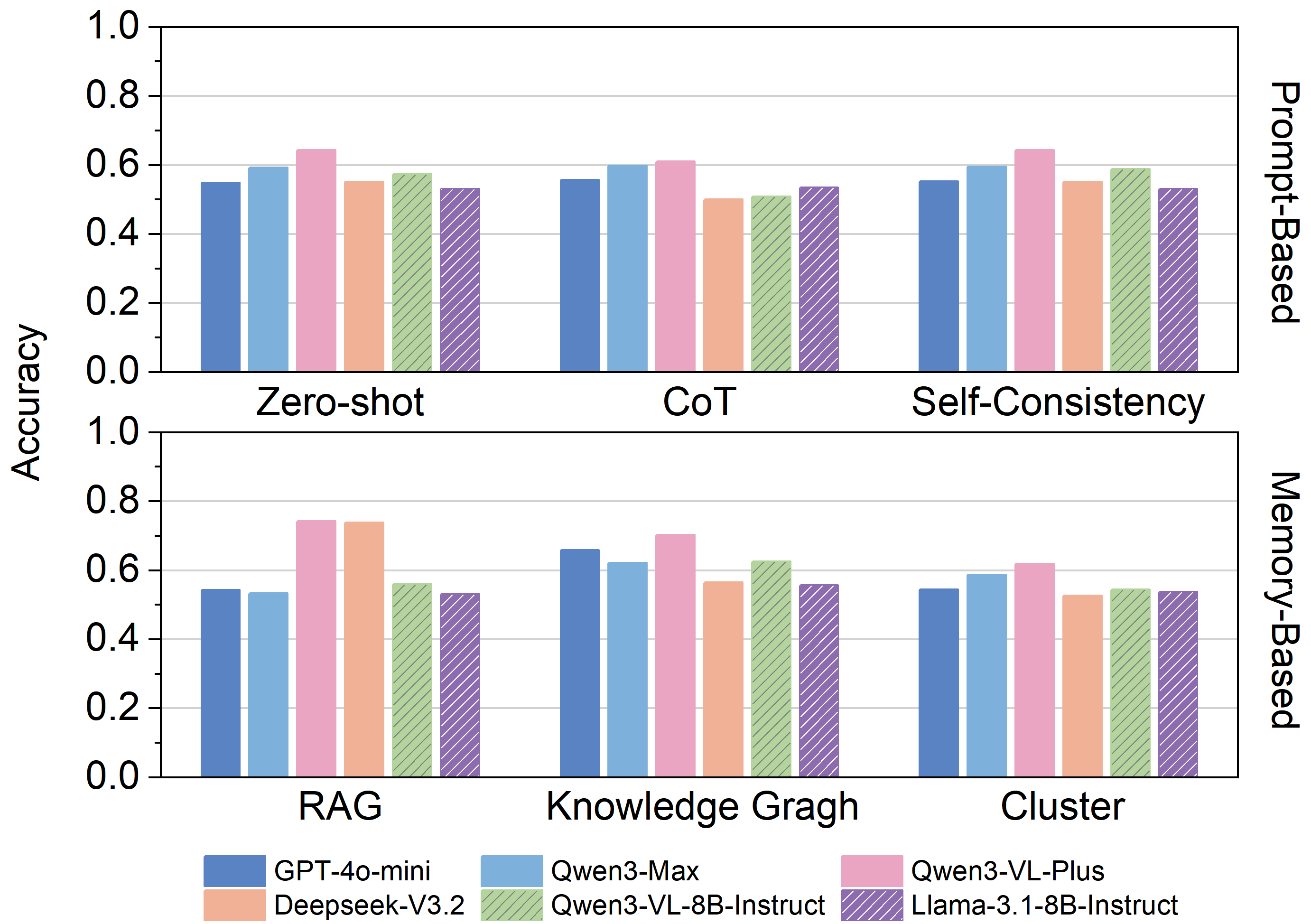}
    \caption{Performance on the \textit{When to Assist} task.}
    \label{fig:accuracy_when}
  \end{subfigure}
  \begin{subfigure}[b]{0.49\columnwidth}
    \centering
    \includegraphics[width=\linewidth]{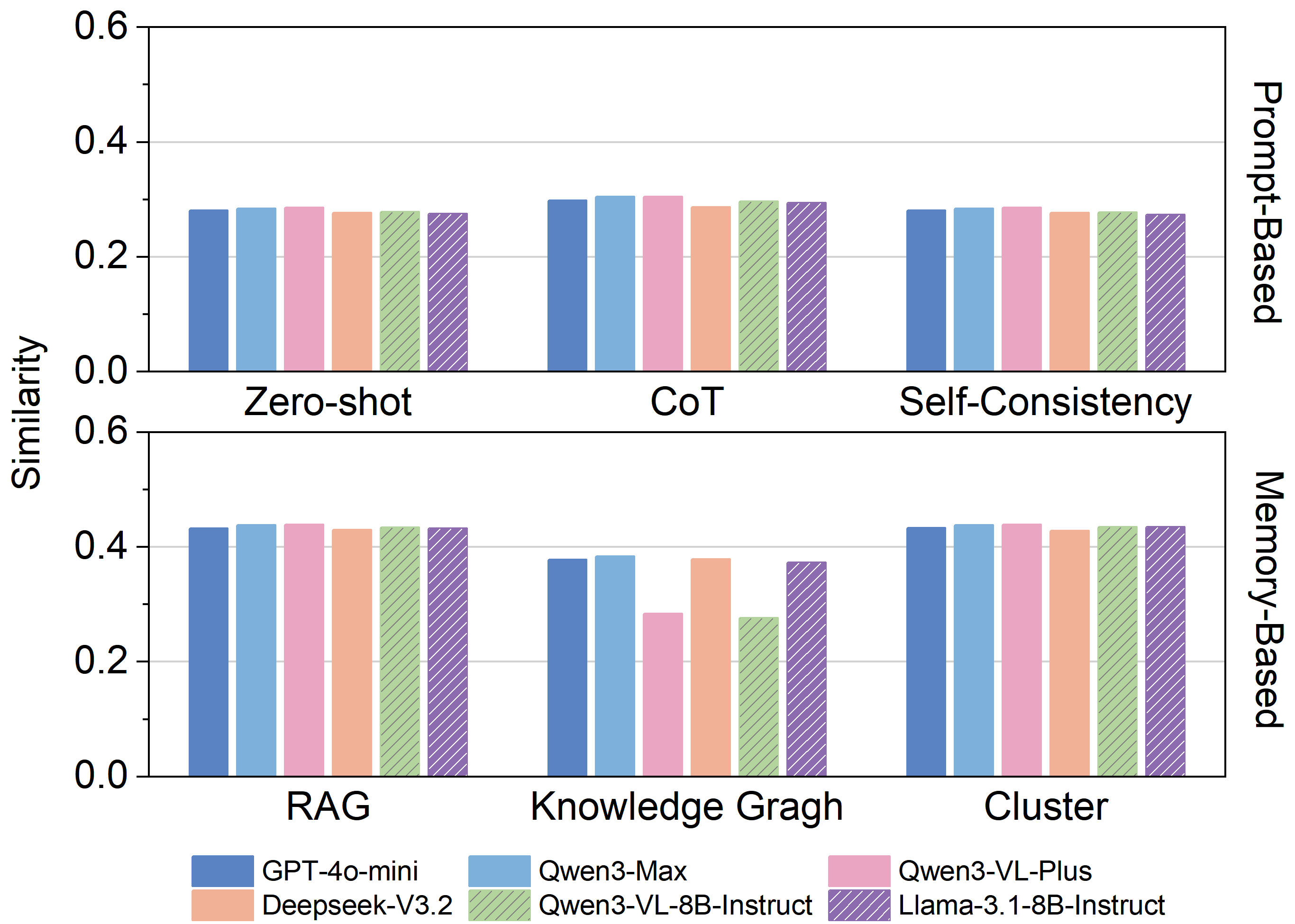}
    \caption{Performance on the \textit{How to Assist} task.}
    \label{fig:similarity_how}
  \end{subfigure}
  \caption{\textbf{Performance comparison of different models and methods.} We evaluate six models using prompt-based methods (Zero-shot, CoT, Self-Consistency) and memory-based methods (RAG, Knowledge Graph, Cluster). (a) Accuracy on timing prediction. (b) Semantic similarity on content prediction.}
  \label{fig:model_performance}
  \vspace{-20pt}
\end{figure}

\vspace{-5pt}
\subsection{Research Question 3: Impact of Real-World Data}
\vspace{-5pt}

A fundamental question in developing proactive assistance systems is whether real-world human data provides unique value compared to synthetic data generated by LLMs. We investigate whether task-specific fine-tuning on real-world data can substantially improve model performance, and how this compares to training on LLM-synthesized data.

We construct two training sets: (1) \textbf{Real-world data}: 741 instances sampled from our collected dataset, comprising diverse user profiles and authentic interaction patterns; (2) \textbf{Synthetic data}: An equivalent number of instances generated following \citet{sun2025ppp}.
\textbf{Fine-tuning Methods.} We employ two parameter-efficient fine-tuning strategies to adapt pre-trained models: (1) \textbf{Supervised Fine-Tuning (SFT)}: Learning rate of $2 \times 10^{-5}$, batch size of 16, for 3 epochs; (2) \textbf{Low-Rank Adaptation (LoRA)}: Fine-tuning with rank $r=16$, learning rate of $2 \times 10^{-4}$, batch size of 16, for 3 epochs.

\begin{table}[t]
  \centering
  \caption{Impact of training data source on open-source models. We compare Zero-shot baseline with models fine-tuned on real-world vs. synthetic data using SFT and LoRA. Abbreviations: Acc.=Accuracy, Int. Acc.=Intention Accuracy, Sem. Sim.=Semantic Similarity. Best results per model are \textbf{bolded}.}
  \label{tab:real_world_data}
  \resizebox{\columnwidth}{!}{
  \begin{tabular}{llcccc}
    \toprule
    \textbf{Method} & \textbf{Data} & \textbf{Acc.} & \textbf{F1} & \textbf{Int. Acc.} & \textbf{Sem. Sim.} \\
    \midrule
    \rowcolor{gray!15} \multicolumn{6}{l}{\textbf{LLaMA-3.1-8B-Instruct}} \\
    Zero-shot & -- & 57.3\% & 66.7\% & 32.3\% & 0.275 \\
    \cmidrule(lr){1-6}
    \multirow{2}{*}{SFT} & Synthetic & 62.1\% & 70.2\% & 34.8\% & 0.312 \\
                         & Real-world & \textbf{74.0\%} & \textbf{78.5\%} & \textbf{42.1\%} & \textbf{0.385} \\
    \cmidrule(lr){1-6}
    \multirow{2}{*}{LoRA} & Synthetic & 60.5\% & 68.9\% & 33.6\% & 0.298 \\
                          & Real-world & 71.2\% & 76.3\% & 40.5\% & 0.372 \\
    \midrule
    \rowcolor{gray!15} \multicolumn{6}{l}{\textbf{Qwen3-VL-8B-Instruct}} \\
    Zero-shot & -- & 51.7\% & 66.1\% & 35.3\% & 0.276 \\
    \cmidrule(lr){1-6}
    \multirow{2}{*}{SFT} & Synthetic & 54.8\% & 67.8\% & 36.2\% & 0.295 \\
                         & Real-world & \textbf{63.5\%} & \textbf{72.4\%} & \textbf{41.8\%} & \textbf{0.358} \\
    \cmidrule(lr){1-6}
    \multirow{2}{*}{LoRA} & Synthetic & 53.2\% & 67.1\% & 35.9\% & 0.288 \\
                          & Real-world & 61.8\% & 71.2\% & 40.2\% & 0.345 \\
    \bottomrule
  \end{tabular}
  }
  \vspace{-20pt}
\end{table}

As shown in Table~\ref{tab:real_world_data}, both fine-tuning methods significantly enhance model performance compared to zero-shot baselines. Notably, for LLaMA-3.1-8B-Instruct, fine-tuning on real-world data leads to substantial improvements, boosting Accuracy from 57.3\% to 74.0\% (+16.7\%). More importantly, models trained on real-world data consistently outperform those trained on synthetic data across all metrics. For instance, LLaMA with SFT on real-world data achieves 74.0\% Accuracy compared to 62.1\% with synthetic data (+11.9\%), and Intention Accuracy improves from 34.8\% to 42.1\% (+7.3\%). This performance gap demonstrates that authentic human interaction patterns contain valuable signals that synthetic data cannot fully replicate. This finding underscores the critical importance of collecting and utilizing real-world human data for developing effective proactive assistance systems.

\section{Conclusion}

In this paper, we present \textbf{ProAgentBench}, the first rigorous benchmark designed to evaluate proactive agents within continuous real-world workflows. Addressing the limitations of synthetic and isolated datasets, we construct a privacy-compliant dataset capturing over 28,000 events from 500+ hours of authentic user activities, preserving critical pre-assistance behavioral patterns. We propose a hierarchical framework to systematically evaluate proactive capabilities in timing prediction and content generation. Our experiments reveal that real-world training data and long-term memory integration are pivotal for agent performance. We believe ProAgentBench provides a solid foundation for advancing the development of context-aware, proactive AI systems that seamlessly integrate into human workflows.

\section{Impact Statements}

\subsection{Limitations}
Our dataset has several limitations that should be acknowledged. First, participant bias exists as our volunteers are limited to specific professions, technology stacks, regions, and languages; differences in OS and application ecosystems may affect the generalization of trained models to broader populations. Second, our 1Hz sampling rate may miss very short interactions, and unstable or dynamically changing window titles can introduce annotation errors. Third, our aggressive privacy filtering pipeline, while essential for ethical data release, may systematically exclude certain interaction patterns involving sensitive content, potentially biasing the dataset toward less privacy-sensitive workflows.

\subsection{Ethics Statement}
All participants in this study were voluntary contributors who were fully informed about the data collection process. Prior to participation, each individual was clearly briefed on: (1) the types of data collected, including screenshots and application metadata; (2) the research purpose and potential academic publication; (3) comprehensive privacy protection measures; and (4) the unconditional right to withdraw at any time with complete data deletion guaranteed within 7 days.

To protect participant privacy, we implemented a rigorous three-stage pipeline: real-time filtering of sensitive applications (e.g., banking, medical), VLM-based automatic detection of personally identifiable information (names, phone numbers, emails, passwords), and mandatory participant review where individuals retained final authority over all retention decisions. Participants could mark any screenshot for deletion without providing justification.

We acknowledge that screen-level behavioral data carries inherent surveillance risks if misused. To mitigate these concerns, we enforce strict access controls: raw screenshots are restricted to approved researchers under signed data use agreements, while public releases contain only de-identified data and aggregated statistics. All usage is governed by a research-only license that explicitly prohibits re-identification attempts, commercial applications, and any form of user monitoring or profiling. We believe these comprehensive safeguards ensure that the scientific benefits of ProAgentBench substantially outweigh potential risks.

\subsection{Broader Impact}
ProAgentBench has both positive and negative potential impacts. On the positive side, it enables research on proactive AI assistants that anticipate user needs, potentially improving productivity and reducing cognitive load in knowledge work. On the negative side, advances in this area may contribute to over-reliance on AI assistance or enable intrusive applications if privacy safeguards are bypassed. We encourage the research community to develop proactive agents that respect user autonomy and provide transparent, controllable assistance.

\subsection{Future Work}
Several directions remain for future exploration. First, incorporating richer sensor modalities such as keyboard dynamics, mouse trajectories, and system-level events could enable finer-grained behavior modeling. Second, developing stronger sequence models capable of capturing long-range temporal dependencies may improve prediction accuracy for complex workflows. Third, conducting user studies with online deployment of proactive assistants would provide valuable insights into real-world usability, user acceptance, and the appropriate balance between proactivity and intrusiveness.
\nocite{*}
\bibliography{example_paper}

@inproceedings{langley00,
 author    = {P. Langley},
 title     = {Crafting Papers on Machine Learning},
 year      = {2000},
 pages     = {1207--1216},
 editor    = {Pat Langley},
 booktitle     = {Proceedings of the 17th International Conference
              on Machine Learning (ICML 2000)},
 address   = {Stanford, CA},
 publisher = {Morgan Kaufmann}
}

@TechReport{mitchell80,
  author = 	 "T. M. Mitchell",
  title = 	 "The Need for Biases in Learning Generalizations",
  institution =  "Computer Science Department, Rutgers University",
  year = 	 "1980",
  address =	 "New Brunswick, MA",
}

@phdthesis{kearns89,
  author = {M. J. Kearns},
  title =  {Computational Complexity of Machine Learning},
  school = {Department of Computer Science, Harvard University},
  year =   {1989}
}

@Book{MachineLearningI,
  editor = 	 "R. S. Michalski and J. G. Carbonell and T.
		  M. Mitchell",
  title = 	 "Machine Learning: An Artificial Intelligence
		  Approach, Vol. I",
  publisher = 	 "Tioga",
  year = 	 "1983",
  address =	 "Palo Alto, CA"
}

@Book{DudaHart2nd,
  author =       "R. O. Duda and P. E. Hart and D. G. Stork",
  title =        "Pattern Classification",
  publisher =    "John Wiley and Sons",
  edition =      "2nd",
  year =         "2000"
}

@InCollection{Newell81,
  author =       "A. Newell and P. S. Rosenbloom",
  title =        "Mechanisms of Skill Acquisition and the Law of
                  Practice", 
  booktitle =    "Cognitive Skills and Their Acquisition",
  pages =        "1--51",
  publisher =    "Lawrence Erlbaum Associates, Inc.",
  year =         "1981",
  editor =       "J. R. Anderson",
  chapter =      "1",
  address =      "Hillsdale, NJ"
}

@Article{Samuel59,
  author = 	 "A. L. Samuel",
  title = 	 "Some Studies in Machine Learning Using the Game of
		  Checkers",
  journal =	 "IBM Journal of Research and Development",
  year =	 "1959",
  volume =	 "3",
  number =	 "3",
  pages =	 "211--229"
}

@article{hong2009context,
  title={Context-aware system for proactive personalized service based on context history},
  author={Hong, Jinyoung and Suh, Eui-Ho and Kim, Junho and Kim, Seon-Young},
  journal={Expert Systems with Applications},
  volume={36},
  number={4},
  pages={7448--7457},
  year={2009}
}

@inproceedings{woerndl2011model,
  title={A model for proactivity in mobile, context-aware recommender systems},
  author={Woerndl, Wolfgang and Huebner, Johannes and Bader, Rolf and Gallego-Vico, Daniel},
  booktitle={Proceedings of the Fifth ACM Conference on Recommender Systems},
  pages={273--276},
  year={2011}
}

@article{meurisch2020exploring,
  title={Exploring user expectations of proactive AI systems},
  author={Meurisch, Christian and Mihale-Wilson, Cristina and Hawlitschek, Anna and Giger, Florian and Hinz, Oliver and Seip, Benedict and M{\"u}hlh{\"a}user, Max},
  journal={Proceedings of the ACM on Interactive, Mobile, Wearable and Ubiquitous Technologies},
  volume={4},
  number={4},
  pages={1--22},
  year={2020}
}

@inproceedings{oh2011probabilistic,
  title={Probabilistic plan recognition for intelligent information agents-towards proactive software assistant agents},
  author={Oh, Jean and Meneguzzi, Felipe and Sycara, Katia},
  booktitle={Proceedings of the International Conference on Agents and Artificial Intelligence},
  pages={281--287},
  year={2011}
}

@inproceedings{jones2024designing,
  title={Designing a Proactive Context-Aware AI Chatbot for People's Long-Term Goals},
  author={Jones, Brent and Xu, Yun and Li, Qingyang and Scherer, Stefan},
  booktitle={Proceedings of the CHI Conference on Human Factors in Computing Systems},
  pages={1--17},
  year={2024}
}

@inproceedings{deka2017rico,
  title={Rico: A mobile app dataset for building data-driven design applications},
  author={Deka, Biplab and Huang, Zifeng and Franzen, Chad and Hibschman, Joshua and Afergan, Daniel and Li, Yang and Nichols, Jeffrey and Kumar, Ranjitha},
  booktitle={Proceedings of the 30th Annual ACM Symposium on User Interface Software and Technology},
  pages={845--854},
  year={2017}
}

@article{deng2023mind2web,
  title={Mind2web: Towards a generalist agent for the web},
  author={Deng, Xiang and Gu, Yu and Zheng, Boyuan and Chen, Shijie and Stevens, Samuel and Wang, Boshi and Sun, Huan and Su, Yu},
  journal={Advances in Neural Information Processing Systems},
  volume={36},
  year={2023}
}

@article{zhou2023webarena,
  title={Webarena: A realistic web environment for building autonomous agents},
  author={Zhou, Shuyan and Xu, Frank F and Zhu, Hao and Zhou, Xuhui and Lo, Robert and Sridhar, Abishek and Cheng, Xianyi and Bisk, Yonatan and Fried, Daniel and Neubig, Graham},
  journal={arXiv preprint arXiv:2307.13854},
  year={2023}
}

@article{gao2023assistgui,
  title={Assistgui: Task-oriented desktop graphical user interface automation},
  author={Gao, Difei and Ji, Lei and Bai, Zechen and Ouyang, Mingyu and Li, Peiran and Mao, Dongxing and Wu, Qinchen and Zhang, Weichen and Wang, Peiyi and Shou, Mike Zheng},
  journal={arXiv preprint arXiv:2312.13108},
  year={2023}
}

@inproceedings{kapoor2024omniact,
  title={Omniact: A dataset and benchmark for enabling multimodal generalist autonomous agents for desktop and web},
  author={Kapoor, Raghav and Butala, Yash Parag and Russak, Melisa and Koh, Joon Sung and Isahagian, Vagner and Muthusamy, Vinod and Khalil, Ihab F and Rizvi, Asim Munawar},
  booktitle={European Conference on Computer Vision},
  pages={161--179},
  year={2024}
}

@article{chen2024gui,
  title={Gui-world: A video benchmark and dataset for multimodal gui-oriented understanding},
  author={Chen, Dongping and Huang, Yue and Wu, Siyuan and Tang, Jingyu and Chen, Liuyi and Bai, Yilin and He, Zhigang and Zhou, Huichi and Sun, Lichao},
  journal={arXiv preprint arXiv:2406.10819},
  year={2024}
}

@article{rawles2024androidworld,
  title={Androidworld: A dynamic benchmarking environment for autonomous agents},
  author={Rawles, Christopher and Clinckemaillie, Sarah and Chang, Yifan and Waltz, Jonathan and Lau, Gabrielle and Fair, Marybeth and Li, Alice and Riva, Oriana},
  journal={arXiv preprint arXiv:2405.14573},
  year={2024}
}

@article{lu2024proactive,
  title={Proactive agent: Shifting llm agents from reactive responses to active assistance},
  author={Lu, Yaxi and Yang, Shenzhi and Qian, Cheng and Chen, Guirong and Luo, Qinyu and Wu, Yesai and Wang, Huadong and Cong, Xin and Zhang, Zhong and Lin, Yankai and Liu, Weiwen and Wang, Yasheng and Liu, Zhiyuan and Liu, Fangming and Sun, Maosong},
  journal={arXiv preprint arXiv:2410.12361},
  year={2024}
}

@article{yang2025contextagent,
  title={ContextAgent: Context-Aware Proactive LLM Agents with Open-World Sensory Perceptions},
  author={Yang, Bufang and Xu, Lixing and Zeng, Lichao and Liu, Kaiyan and Jiang, Shiyi and Lu, Weiwen and Cong, Xin and Lu, Yaxi and Lin, Yankai and Sun, Maosong},
  journal={arXiv preprint arXiv:2505.14668},
  note={Accepted by NeurIPS 2025},
  year={2025}
}

@article{yang2025fingertip,
  title={Fingertip 20k: A benchmark for proactive and personalized mobile llm agents},
  author={Yang, Qinglong and Li, Haixu and Zhao, Haocheng and Yan, Xiaofan and Ding, Jianfei and Xu, Fei and Han, Zhengyu and Pan, Lingyu and Cao, Yuntao and Shi, Yuanchun},
  journal={arXiv preprint arXiv:2507.21071},
  year={2025}
}

@article{liu2025proactiveeval,
  title={ProactiveEval: A Unified Evaluation Framework for Proactive Dialogue Agents},
  author={Liu, Tianjian and Wan, Fengzhe and Guo, Jiacheng and Quan, Xiaojun},
  journal={arXiv preprint arXiv:2508.20973},
  year={2025}
}

@article{sun2025ppp,
  title={Training Proactive and Personalized LLM Agents},
  author={Sun, Weiwei and Zhou, Xuhui and Du, Weihua and Wang, Xingyao and Welleck, Sean and Neubig, Graham and Sap, Maarten and Yang, Yiming},
  journal={arXiv preprint arXiv:2511.02208},
  year={2025},
  note={Carnegie Mellon University}
}

@inproceedings{salemi-etal-2024-lamp,
    title = "{L}a{MP}: When Large Language Models Meet Personalization",
    author = "Salemi, Alireza  and
      Mysore, Sheshera  and
      Bendersky, Michael  and
      Zamani, Hamed",
    editor = "Ku, Lun-Wei  and
      Martins, Andre  and
      Srikumar, Vivek",
    booktitle = "Proceedings of the 62nd Annual Meeting of the Association for Computational Linguistics (Volume 1: Long Papers)",
    month = aug,
    year = "2024",
    address = "Bangkok, Thailand",
    publisher = "Association for Computational Linguistics",
    url = "https://aclanthology.org/2024.acl-long.399/",
    doi = "10.18653/v1/2024.acl-long.399",
    pages = "7370--7392"
}

@inproceedings{tan-etal-2024-democratizing,
    title = "Democratizing Large Language Models via Personalized Parameter-Efficient Fine-tuning",
    author = "Tan, Zhaoxuan  and
      Zeng, Qingkai  and
      Tian, Yijun  and
      Liu, Zheyuan  and
      Yin, Bing  and
      Jiang, Meng",
    editor = "Al-Onaizan, Yaser  and
      Bansal, Mohit  and
      Chen, Yun-Nung",
    booktitle = "Proceedings of the 2024 Conference on Empirical Methods in Natural Language Processing",
    month = nov,
    year = "2024",
    address = "Miami, Florida, USA",
    publisher = "Association for Computational Linguistics",
    url = "https://aclanthology.org/2024.emnlp-main.372/",
    doi = "10.18653/v1/2024.emnlp-main.372",
    pages = "6476--6491"
}

@misc{richardson2023personalization,
      title={Integrating Summarization and Retrieval for Enhanced Personalization via Large Language Models}, 
      author={Chris Richardson and Yao Zhang and Kellen Gillespie and Sudipta Kar and Arshdeep Singh and Zeynab Raeesy and Omar Zia Khan and Abhinav Sethy},
      year={2023},
      eprint={2310.20081},
      archivePrefix={arXiv},
      primaryClass={cs.CL},
      url={https://arxiv.org/abs/2310.20081}, 
}

@inproceedings{
liu2025mind,
title={Mind Your Step (by Step): Chain-of-Thought can Reduce Performance on Tasks where Thinking Makes Humans Worse},
author={Ryan Liu and Jiayi Geng and Addison J. Wu and Ilia Sucholutsky and Tania Lombrozo and Thomas L. Griffiths},
booktitle={Forty-second International Conference on Machine Learning},
year={2025},
url={https://openreview.net/forum?id=J3gzdbYZxS}
}

@article{
zheng2025curse,
title={The Curse of CoT: On the Limitations of Chain-of-Thought in In-Context Learning},
author={Tianshi Zheng and Yixiang Chen and Chengxi Li and Chunyang Li and Qing Zong and Haochen Shi and Baixuan Xu and Yangqiu Song and Ginny Wong and Simon See},
journal={Transactions on Machine Learning Research},
issn={2835-8856},
year={2025},
url={https://openreview.net/forum?id=7SIrvcYNYj},
note={}
}

@article{goh2008burstiness,
  title={Burstiness and memory in complex systems},
  author={Goh, K-I and Barab{\'a}si, A-L},
  journal={Europhysics Letters},
  volume={81},
  number={4},
  pages={48002},
  year={2008},
  publisher={IOP Publishing}
}

@inproceedings{mark2008cost,
  title={The cost of interrupted work: more speed and stress},
  author={Mark, Gloria and Gudith, Daniela and Klocke, Ulrich},
  booktitle={Proceedings of the SIGCHI Conference on Human Factors in Computing Systems},
  pages={107--110},
  year={2008},
  organization={ACM}
}

@inproceedings{adamczyk2004interruptions,
  title={If not now, when? The effects of interruption at different moments within task execution},
  author={Adamczyk, Piotr D and Bailey, Brian P},
  booktitle={Proceedings of the SIGCHI Conference on Human Factors in Computing Systems},
  pages={271--278},
  year={2004},
  organization={ACM}
}

@inproceedings{iqbal2007disruption,
  title={Disruption and recovery of computing tasks: field study, analysis, and directions},
  author={Iqbal, Shamsi T and Horvitz, Eric},
  booktitle={Proceedings of the SIGCHI Conference on Human Factors in Computing Systems},
  pages={677--686},
  year={2007},
  organization={ACM}
}

@article{czerwinski2004diary,
  title={A diary study of task switching and interruptions},
  author={Czerwinski, Mary and Horvitz, Eric and Wilhite, Susan},
  journal={Proceedings of the SIGCHI Conference on Human Factors in Computing Systems},
  pages={175--182},
  year={2004}
}

@article{parnin2013programmer,
  title={Programmer information needs after memory failure},
  author={Parnin, Chris and DeLine, Robert},
  journal={Proceedings of the IEEE International Conference on Program Comprehension},
  pages={123--132},
  year={2013}
}

@article{alertfatigue2023,
  title={Alert fatigue: A growing challenge in healthcare and technology},
  author={Cash, John J},
  journal={American Journal of Health-System Pharmacy},
  volume={66},
  number={23},
  pages={2098--2101},
  year={2009}
}

@article{proactive_bise2024,
  title={When AI-Based Agents Are Proactive: Implications for Competence and System Satisfaction in Human--AI Collaboration},
  author={Herm, Lukas-Valentin and Steinbach, Theresa and Wanner, Jonas and Janiesch, Christian},
  journal={Business \& Information Systems Engineering},
  year={2024},
  publisher={Springer}
}

@inproceedings{wei2022chain,
  title={Chain-of-thought prompting elicits reasoning in large language models},
  author={Wei, Jason and Wang, Xuezhi and Schuurmans, Dale and Bosma, Maarten and Ichter, Brian and Xia, Fei and Chi, Ed and Le, Quoc and Zhou, Denny},
  booktitle={Advances in Neural Information Processing Systems},
  volume={35},
  pages={24824--24837},
  year={2022}
}

@inproceedings{wang2023selfconsistency,
  title={Self-consistency improves chain of thought reasoning in language models},
  author={Wang, Xuezhi and Wei, Jason and Schuurmans, Dale and Le, Quoc and Chi, Ed and Narang, Sharan and Chowdhery, Aakanksha and Zhou, Denny},
  booktitle={International Conference on Learning Representations},
  year={2023}
}

@article{openai2024gpt4o,
  title={GPT-4o System Card},
  author={{OpenAI}},
  journal={arXiv preprint arXiv:2410.21276},
  year={2024}
}

@article{wang2024qwen2vl,
  title={Qwen2-VL: Enhancing Vision-Language Model's Perception of the World at Any Resolution},
  author={Wang, Peng and Bai, Shuai and Tan, Sinan and Wang, Shijie and Fan, Zhihao and Bai, Jinze and Chen, Keqin and Liu, Xuejing and Wang, Jialin and Ge, Wenbin and others},
  journal={arXiv preprint arXiv:2409.12191},
  year={2024}
}

@misc{qwen3,
  title={Qwen3 Technical Report},
  author={{Qwen Team}},
  year={2025},
  howpublished={\url{https://qwenlm.github.io/blog/qwen3/}}
}

@article{deepseekv3,
  title={DeepSeek-V3 Technical Report},
  author={{DeepSeek-AI}},
  journal={arXiv preprint arXiv:2412.19437},
  year={2024}
}

@article{llama3,
  title={The Llama 3 Herd of Models},
  author={Grattafiori, Aaron and Dubey, Abhimanyu and Jauhri, Abhinav and others},
  journal={arXiv preprint arXiv:2407.21783},
  year={2024}
}

@inproceedings{personax2025,
  title={PersonaX: A Recommendation Agent-Oriented User Modeling Framework for Long Behavior Sequence},
  author={Li, Yunxiao and Wang, Jiangxia and Zhao, Hang and Zhang, Shuo and Liang, Yuyang and Tang, Jing and He, Xiangnan},
  booktitle={Findings of the Association for Computational Linguistics: ACL 2025},
  year={2025}
}

@inproceedings{Lewis2020,
  title     = {Retrieval-Augmented Generation for Knowledge-Intensive {NLP} Tasks},
  author    = {Lewis, Patrick and Perez, Ethan and Piktus, Aleksandra and Petroni, Fabio and Karpukhin, Vladimir and Goyal, Naman and K\"{u}ttler, Heinrich and Lewis, Mike and Yih, Wen-tau and Rockt\"{a}schel, Tim and Riedel, Sebastian and Kiela, Douwe},
  booktitle = {Advances in Neural Information Processing Systems},
  volume    = {33},
  pages     = {9459--9474},
  year      = {2020},
  publisher = {Curran Associates, Inc.}
}

@article{Huang2024,
  title     = {A Survey on Retrieval-Augmented Text Generation for Large Language Models},
  author    = {Huang, Yizheng and Huang, Jimmy},
  journal   = {arXiv preprint arXiv:2404.10981},
  year      = {2024}
}

@inproceedings{Shi2025,
  title     = {{PersonaX}: A Recommendation Agent-Oriented User Modeling Framework for Long Behavior Sequence},
  author    = {Shi, Yunxiao and Xu, Wujiang and Zhang, Zeqi and Zi, Xing and Wu, Qiang and Xu, Min},
  booktitle = {Findings of the Association for Computational Linguistics: ACL 2025},
  pages     = {4362--4378},
  year      = {2025},
  publisher = {Association for Computational Linguistics}
}

@article{dubey2024llama,
  title={The llama 3 herd of models},
  author={Dubey, Abhimanyu and Jauhri, Abhinav and Pandey, Abhinav and Kadian, Abhishek and Al-Dahle, Ahmad and Letman, Aiesha and Mathur, Akhil and Schelten, Alan and Yang, Amy and Fan, Angela and others},
  journal={arXiv e-prints},
  pages={arXiv--2407},
  year={2024}
}

@article{team2024gemini,
  title={Gemini 1.5: Unlocking multimodal understanding across millions of tokens of context},
  author={Team, Gemini and Georgiev, Petko and Lei, Ving Ian and Burnell, Ryan and Bai, Libin and Gulati, Anmol and Tanzer, Garrett and Vincent, Damien and Pan, Zhufeng and Wang, Shibo and others},
  journal={arXiv preprint arXiv:2403.05530},
  year={2024}
}

@article{team2025kimi,
  title={Kimi k2: Open agentic intelligence},
  author={Team, Kimi and Bai, Yifan and Bao, Yiping and Chen, Guanduo and Chen, Jiahao and Chen, Ningxin and Chen, Ruijue and Chen, Yanru and Chen, Yuankun and Chen, Yutian and others},
  journal={arXiv preprint arXiv:2507.20534},
  year={2025}
}
\bibliographystyle{icml2026}

\newpage
\appendix
\onecolumn

\section{Data Release \& Usage}

The dataset will be released in tiered access levels to balance research utility with privacy protection. Raw screenshots are restricted to approved researchers under strict data use agreements. The public release includes de-identified screenshots, derived features, event-level summaries, and evaluation protocols.

All usage is governed by a research-only license that explicitly prohibits re-identification attempts and commercial applications. We provide privacy-minimizing training and evaluation guidelines, along with reproducible scripts and baseline implementations to facilitate adoption by the research community.

\subsection{Data fields}
To support reproducible analyses and downstream modeling, we release the dataset in fully parseable, structured formats, consisting of (i) a SQLite database that stores the core logs and (ii) external annotation/curation files (CSV/JSON/JSONL) that provide traceable labeling and filtering decisions. The database is organized around two primary tables, including \texttt{screenshots} and \texttt{events}, which are linked through a consistent key (\texttt{event\_id}), while the external files are keyed by the event identifier (e.g., \texttt{user} + \texttt{event\_id}) to enable deterministic alignment with database entries.

\noindent\textbf{Screenshot record fields.} The \texttt{screenshots} table provides the record-level information required to locate a screenshot, align it on the timeline, and assess its integrity. It includes the screenshot file name (\texttt{file\_path}) and creation time (\texttt{created\_at}), along with aligned foreground context (\texttt{app\_name}, \texttt{window\_title}). In addition, we store file-level attributes such as content hash (\texttt{file\_hash}), file size (\texttt{file\_size}), and image dimensions (\texttt{width}, \texttt{height}). These attributes enable systematic reporting of data quality (e.g., missing files, duplicates, and abnormal file properties) and provide deterministic signals for integrity checks. Because some \texttt{file\_path} values preserve capture-side absolute paths, reproducible usage typically resolves screenshots by filename and matches them to the local screenshot directory.

\noindent\textbf{Event fields.} The \texttt{events} table provides the event-level representation. It includes an event identifier (\texttt{id}), temporal boundaries (\texttt{start\_time}, \texttt{end\_time}), and event-level context (\texttt{app\_name}, \texttt{window\_title}). Events also contain an LLM-related flag (\texttt{is\_llm\_event}) and textual descriptors (e.g., \texttt{event\_summary}, \texttt{detailed\_description}, and optionally model-generated titles/summaries). In addition, events store a structured conversation field (\texttt{conversation}) as a JSON string, which captures observable interaction cues such as extracted user queries and model responses (e.g., \texttt{user\_queries}, \texttt{llm\_responses}, and \texttt{full\_conversation}). These event fields support platform inference, semantic categorization, and topic representation based on visible input content.

\noindent\textbf{External annotations and curation artifacts.} In addition to database-native fields, we provide external files that make labeling and dataset curation explicit and auditable. These artifacts include (1) recheck/exclusion lists that specify which candidate events should be removed and why (e.g., verified non-LLM cases or unusable entries), (2) intent annotation outputs (e.g., \texttt{user\_intention} with confidence and optional rationales) stored in JSONL/JSON for deterministic replay, and (3) when available, platform tags (e.g., \texttt{llm\_platform}) from verification pipelines that improve the stability of platform attribution beyond app/window heuristics. All external artifacts are indexed by event keys and can be joined back to the database unambiguously, ensuring that the final analysis set and its labels are fully traceable and reproducible.

Overall, the combination of database fields and external annotation/curation files provides a complete, parseable interface to the dataset: the former captures the core screenshot and event logs, while the latter records reproducible labeling and filtering decisions required to construct the analysis-ready subset used in this work.
\subsection{Data organization and file structure}
The dataset is organized with participant (user) as the top-level key. For each participant, we provide both the screenshot files and the structured metadata required to parse and align the logs.

At the file-system level, each participant directory contains a \texttt{screenshots/} folder that stores the screenshots sampled at approximately 1~Hz. Screenshot filenames encode date and time information, which supports convenient retrieval at the day/hour granularity when needed. Some participant folders also include additional pipeline-generated subdirectories (e.g., privacy-processing artifacts), which may contain copies of screenshots or related intermediate outputs.

At the metadata level, each participant directory includes a SQLite database file named \texttt{lifetrace\_privacy\_processed.db}. The database is centered around two core tables, \texttt{screenshots} and \texttt{events}: the \texttt{screenshots} table stores screenshot-level records (timestamps, foreground app, window title, and file attributes), and the \texttt{events} table stores event-level records (event boundaries, context, and semantic fields). These tables are linked via \texttt{screenshots.event\_id} and \texttt{events.id}, allowing each event to be mapped to its associated screenshot sequence and each screenshot to be traced back to its parent event.

In addition, some semantic annotations (e.g., event intent labels) are stored as separate JSON/JSONL files. These files are indexed by event keys (participant/user + \texttt{event\_id}), enabling straightforward alignment with the event records in the SQLite databases.

\subsection{Intention Categories}
\label{app:intention_categories}

To characterize usage intent, we categorize LLM interactions into multiple scenario types based on event semantics. Table~\ref{tab:intention_distribution} presents the complete distribution of user intentions. Overall, information-seeking needs dominate: information lookup and knowledge Q\&A together account for over 55\% of all LLM events. Meanwhile, productive and analytical tasks also represent a substantial share, including data analysis, coding/programming, and content generation. The remaining categories form a long-tail distribution.

\begin{table}[h]
\centering
\caption{Distribution of user intentions across LLM interaction events.}
\label{tab:intention_distribution}
\begin{tabular}{lrr}
\toprule
\textbf{Intention Category} & \textbf{Percentage} & \textbf{Count} \\
\midrule
Information Lookup & 35.10\% & 2,535 \\
Knowledge Q\&A & 20.42\% & 1,475 \\
Data Analysis & 9.17\% & 662 \\
Coding/Programming & 8.72\% & 630 \\
Content Generation & 6.94\% & 501 \\
Advice/Consultation & 5.23\% & 378 \\
Summarization & 4.74\% & 342 \\
Document Editing & 3.28\% & 237 \\
Comparison/Evaluation & 2.17\% & 157 \\
Translation & 1.27\% & 92 \\
Creative Design & 1.16\% & 84 \\
Format Conversion & 0.79\% & 57 \\
Mathematical Calculation & 0.40\% & 29 \\
Uncategorized & 0.55\% & 40 \\
Functional Testing & 0.03\% & 2 \\
Error Correction & 0.01\% & 1 \\
\midrule
\textbf{Total} & \textbf{100.00\%} & \textbf{7,222} \\
\bottomrule
\end{tabular}
\end{table}

\section{Data Collection, Privacy Protection, and Automatic Annotation}
\label{app:data_collection}
We employ LifeTrace\footnote{Project available at \url{https://github.com/FreeU-group/LifeTrace}} to collect volunteers' computer usage behavior data. Through a multi-stage privacy protection process and an event-based data annotation workflow, a high-quality dataset containing Large Language Model (LLM) interaction scenarios is constructed. The entire pipeline consists of three main phases: data collection, privacy protection, and automatic annotation, ensuring the authenticity, privacy security, and annotation accuracy of the data, as is illustrated in Figure~\ref{fig:data-collection-pipeline}.
\subsection{Participants and Setup}
We recruited a diverse group of participants spanning various age ranges and LLM usage frequencies, covering multiple professional scenarios. All participants were required to use devices running Windows 10+ or macOS 12+ to ensure compatibility with our data collection tools. The data collection was conducted over a month, during which participants had full control over the process. We implemented strict consent and withdrawal protocols, allowing participants to pause or exit the study at any time and request the deletion of their data, ensuring ethical compliance and user privacy.

\subsection{Data Collection and Instrumentation}
We use LifeTrace application to collect real-world computer usage data, which monitors user activities with the user's informed consent. LifeTrace collects two types of data: (1) User screen screenshots captured at a rate of 1Hz; (2) Application usage logs, recording timestamps, application names, and window titles. Based on application switching and temporal continuity, LifeTrace automatically segments continuous user activities into discrete events. Each event represents a complete interaction period of the user in a specific application and window environment.

\subsection{Quality Filtering}
We implement multiple filtering mechanisms to improve dataset quality. 
For quality filtering, the following criteria are applied: (1) Events with a duration of $<3$ seconds are excluded, as they typically lack meaningful LLM interactions; (2) Events with an abnormally long duration ($>1$ hour) are reviewed to rule out potential system anomalies; (3) Each event must be associated with at least one valid screenshot; (4) For LLM events, we prioritize retaining events with $\geq 3$ screenshots to ensure complete conversation context; (5) We verify conversation records of LLM events, ensuring the conversation records is not empty.

The system integrates fault-tolerance mechanisms, including automatic API retries, JSON parsing error recovery, and default annotations when VLM fails. SHA-256 file hashing is used for screenshot deduplication to eliminate redundant data. This comprehensive quality control process yields a high-quality dataset with an annotation success rate of 97.6\% (verified through manual inspection of 100 randomly sampled events).

\subsection{Privacy Protection}
\label{app:privacy}
Prior to annotation, a three-stage privacy protection process integrating automatic detection, manual verification, and rule-based filtering is implemented to ensure comprehensive privacy safeguards.

\paragraph{Phase 1: VLM-based Preliminary Judgment}
Qwen3-VL-Plus~\citep{qwen3} is applied to perform multimodal privacy detection on all screenshots as the first line of defense. The model analyzes visual content and OCR-extracted text to identify sensitive information, including names, phone numbers, email addresses, ID card numbers, bank card numbers, passwords, and facial images. For each screenshot, the VLM generates: (1) Privacy risk level (safe/moderate/high-risk); (2) Type of detected privacy information; (3) Recommended action (retain/blur/delete); (4) Scene description for potential replacement. This automated phase provides a high-recall preliminary screening to capture possible sensitive content.

\paragraph{Phase 2: Volunteer Correction}
To address the limitations of automatic detection, volunteers review screenshots marked as safe/moderate and recommended for retention by the VLM, and make a final decision for each image: retain, blur, or delete. This human-in-the-loop approach ensures that context-sensitive information missed by the VLM can be identified, and participants retain control over their own privacy boundaries. We provide volunteers with clear data review guidelines and examples to ensure their fully informed consent regarding data upload.

\paragraph{Phase 3: Rule-based Filtering}
After manual verification, a rule-based system conducts final validation to capture edge cases and enforce consistency. This phase applies deterministic rules, including: (1) Pattern matching for common privacy identifiers (regular expressions for phone numbers, email formats, and ID card numbers); (2) File metadata checks (e.g., screenshots with window titles containing specific keywords are flagged); (3) Consistency verification (e.g., if multiple screenshots in the same event are deleted, adjacent screenshots will be re-evaluated); (4) OCR text cleaning using predefined replacement patterns.
Screenshots that pass all three phases are migrated to a new database with additional privacy-related metadata fields, while high-risk screenshots are permanently deleted and replaced with scene descriptions. Critically, the original OCR text is deleted to prevent privacy leakage.

\subsection{Automatic LLM Event Annotation}
\label{app:annotation}
We developed an event-level automatic annotation process to identify LLM usage scenarios. Unlike traditional screenshot-level classification, our method operates at the event level to leverage temporal context across multiple screenshots.

For each event, we first sample up to 6 screenshots (the first 3 and the last 3) to balance computational cost and information retention. Deleted screenshots are replaced with scene descriptions. We extract the first 500 characters from the OCR results of each screenshot and organize them in chronological order. These multi-modal inputs (images, OCR text, and event metadata including application name, window title, and duration) are fed into Qwen3-VL-Plus via a carefully designed prompt. Our prompt explicitly instructs the model to: (1) Determine whether the event represents an LLM usage scenario; (2) If positive, identify the LLM platform (ChatGPT/Claude/Cursor, etc.) and interaction type (text conversation/code generation/image generation/multi-modal); (3) Generate a concise event summary ($\leq 20$ words); (4) Extract the complete conversation, including all user queries and LLM responses. The model output is constrained to JSON format for structured data extraction. The VLM identifies LLM usage through multiple features: (a) Visual cues, including chat interface layouts, brand logos (ChatGPT icon, Claude logo), and UI components (message bubbles, send buttons); (b) Text patterns in OCR results, such as alternating questions and answers, platform identifiers ("ChatGPT says:", "Claude:"), and generation markers (code blocks, mathematical formulas); (c) Metadata signals, including application names (cursor.exe, chrome.exe), window titles containing LLM platform names, and typical interaction durations ($>30$ seconds).

\section{Memory-Based Methods implementation Details}

\subsection{Retrieval-Augmented Memory}
\label{sec:rag_memory}

While the Knowledge Graph captures aggregated semantic priors, we implement a Retrieval-Augmented Generation (RAG) module to provide LLMs with user-specific historical context at inference time. This approach retrieves concrete historical events similar to the current context and injects them verbatim into the prompt as external memory. This setting follows standard RAG-based memory augmentation paradigms~\cite{Huang2024, Lewis2020}, adapted here with strict temporal constraints.

\paragraph{Memory Construction.}
For each user $u$, we define their episodic memory $\mathcal{M}_u$ derived from their specific subset of the training data $\mathcal{D}_u = \{(a_i, w_i, u_i, h_i, y_i, t_i) \in \mathcal{D} \mid u_i = u\}$, where $t_i$ denotes the timestamp. 
We serialize each interaction record into a structured textual document $d_i$ via a template $T(a_i, w_i, y_i)$, encompassing the active application $a_i$, window title $w_i$, and the ground-truth intent $y_i$. 
We then employ a fixed embedding model $\phi(\cdot)$ to map each document to a dense vector space:
\begin{equation}
    \mathbf{v}_i = \phi(d_i), \quad \forall i \in \mathcal{D}_u
\end{equation}
The resulting memory store $\mathcal{M}_u = \{(\mathbf{v}_i, d_i)\}_{i=1}^{|\mathcal{D}_u|}$ acts as a key-value index, constructed exclusively from training data to ensure strict user isolation.

\paragraph{Temporal-Constrained Retrieval.}
Given a query context $x = (a, w, u, h, t)$, we generate a query embedding $\mathbf{q} = \phi(T(a, w, \emptyset))$. To retrieve relevant context without violating causality, we enforce a strict temporal constraint ensuring only past events are accessible. The retrieval set $\mathcal{R}$ consists of the top-$5$ neighbors based on cosine similarity:
\begin{equation}
    \mathcal{R} = \mathop{\text{arg} \quad {top-}5}_{(\mathbf{v}_i, d_i) \in \mathcal{M}_u} \left( \frac{\mathbf{q} \cdot \mathbf{v}_i}{\|\mathbf{q}\| \|\mathbf{v}_i\|} \right) \quad \text{s.t.} \quad t_i < t
\end{equation}
This mechanism effectively filters out future information, simulating a realistic setting where the agent only has access to the user's history up to the present moment.

\paragraph{Prompt Augmentation.}
Each retrieved memory item is formatted as a \textit{Memory Block} containing its application name, window title, and concise semantic descriptions. The retrieved memory blocks are concatenated to form the memory context $\mathcal{C}_{\text{RAG}}$, which is injected into the prompt alongside the task description:
\begin{equation}
    \tilde{x} = x_{\text{context}} \oplus [\texttt{MEMORY}: \mathcal{C}_{\text{RAG}}] \oplus x_{\text{task}}
\end{equation}
$\mathcal{C}_{\text{RAG}}$ presents the LLM with full narrative examples (e.g., \textit{``In a similar context with VSCode, the user previously searched for generic syntax help''}). This allows the model to leverage few-shot in-context learning to refine its intent understanding based on precedent.

\paragraph{Complexity.}
Memory construction incurs $O(N)$ computational cost for a single pass of offline embedding. During inference, retrieval operates in $O(\log |\mathcal{D}_u|)$ time using Approximate Nearest Neighbor indexing, with end-to-end retrieval latency below $10$\,ms per query in practice. The method introduces no additional trainable parameters and incurs negligible cost relative to standard LLM inference.

\subsection{Knowledge Graph Memory Augmentation}
\label{sec:kg_memory}

To capture user behavioral patterns across applications, we construct a lightweight knowledge graph (KG) from historical interaction data and use it to augment inference-time prompts with contextual priors. This memory-augmented approach follows recent work on personalized LLMs~\cite{salemi-etal-2024-lamp,tan-etal-2024-democratizing}, where user-specific information is retrieved and prepended to prompts without model fine-tuning.

\paragraph{Graph Construction.}
Given a training set $\mathcal{D} = \{(a_i, w_i, u_i, h_i, y_i)\}_{i=1}^{N}$, where $a_i$ denotes the active application (e.g., Chrome, VSCode, Word), $w_i$ the window title, $u_i$ the user identifier, $h_i = [a_{i}^{(1)}, \ldots, a_{i}^{(k)}]$ the recent application history, and $y_i = (b_i, c_i)$ the ground-truth labels for help-needed (binary) and intention (categorical), we construct a heterogeneous graph $\mathcal{G} = (\mathcal{V}, \mathcal{E})$ with four node types: \textsc{App} (application software), \textsc{Keyword} (window title tokens), \textsc{Intent} (intention categories), and \textsc{User}.

For each application $a \in \mathcal{A}$, we compute empirical priors:
\begin{align}
    P_{\text{help}}(a) &= \frac{|\{i : a_i = a \land b_i = \texttt{True}\}|}{|\{i : a_i = a\}|} \\
    P_{\text{intent}}(c \mid a) &= \frac{|\{i : a_i = a \land c_i = c \land b_i = \texttt{True}\}|}{|\{i : a_i = a \land b_i = \texttt{True}\}|}
\end{align}

Additionally, we extract keywords from window titles using tokenization and stopword filtering, then compute keyword-conditioned intent probabilities:
\begin{equation}
    P_{\text{intent}}(c \mid k) = \frac{|\{i : k \in \text{keywords}(w_i) \land c_i = c\}|}{|\{i : k \in \text{keywords}(w_i)\}|}
\end{equation}
where $\text{keywords}(\cdot)$ extracts the top-5 informative tokens from each window title.

We also track application transition patterns from the history sequence, adding directed \textsc{Transition} edges between consecutive applications with frequency-based weights.

\paragraph{Context Retrieval.}
At inference, given a test sample $(a, w, u, h)$, we query $\mathcal{G}$ to retrieve a context tuple:
\begin{equation}
    \mathcal{C}(a, w, h) = \Big( P_{\text{help}}(a), \; \Pi_a, \; \Pi_w \Big)
\end{equation}
where $\Pi_a = \{(c, P_{\text{intent}}(c \mid a)) : c \in \mathcal{I}_a\}$ denotes app-based intent priors, and $\Pi_w = \{(c, \bar{P}_{\text{intent}}(c \mid w))\}$ aggregates keyword-based priors by averaging over extracted keywords:
\begin{equation}
    \bar{P}_{\text{intent}}(c \mid w) = \frac{1}{|K_w|} \sum_{k \in K_w} P_{\text{intent}}(c \mid k)
\end{equation}
We retain only intents exceeding a frequency threshold $\tau = 0.05$ for app-based priors and $\tau = 0.1$ for keyword-based priors to reduce noise.

\paragraph{Prompt Augmentation.}
The retrieved context $\mathcal{C}$ is serialized into natural language and inserted before the task instruction in the prompt:
\begin{equation}
    \tilde{x} = x_{\text{context}} \oplus [\texttt{MEMORY}: \mathcal{C}] \oplus x_{\text{task}}
\end{equation}
The memory section presents distributional hints in interpretable form (e.g., ``\textit{Based on historical patterns for this application: Code Programming 45\%, Knowledge Q\&A 30\%}''). This provides the model with empirical priors as soft guidance without constraining its predictions, allowing it to override historical patterns when current context suggests deviation~\cite{richardson2023personalization}.

\paragraph{Complexity.}
Graph construction requires a single pass over $\mathcal{D}$ with $O(N)$ time and $O(|\mathcal{A}| \cdot |\mathcal{I}| + |\mathcal{K}| \cdot |\mathcal{I}|)$ space, where $|\mathcal{K}|$ is the keyword vocabulary size. Inference-time retrieval operates in $O(|K_w|)$ for keyword lookup, adding negligible overhead ($<$1ms per query), making the approach suitable for real-time proactive assistance.

\subsection{Cluster-Based Persona Memory}
\label{sec:cluster_memory}

While the Knowledge Graph captures statistical priors and RAG retrieves raw episodes, we implement a Cluster-Based Persona memory that summarizes a user’s historical behaviors into a compact set of natural language personas. This approach first groups historical interactions into coherent behavior clusters and then uses a large language model to generate high-level textual descriptions for each cluster. The resulting descriptions serve as user-level behavioral personas and are injected into the prompt at inference time. Both the clustering and sampling procedures strictly follow the PersonaX protocol~\cite{Shi2025}.

\paragraph{Hierarchical Behavior Clustering.}
For each user $u$, we construct the persona set exclusively from the training history $\mathcal{D}_u$. To strictly prevent data leakage, all events occurring in the evaluation period are removed prior to construction.
Each remaining event $e_i$ is serialized into text (by concatenating its application name, window title, event summary and detailed description) before being embedded into a dense vector $v_i$ using a fixed embedding model. We then perform hierarchical clustering over the event embeddings, following the PersonaX protocol without modification. Events are clustered separately for LLM-related and non-LLM-related activities, with a distance threshold of $0.7$ and a maximum of $15$ clusters per category.

\paragraph{Prototypical-Diverse Sampling.}
To summarize each cluster $C_j$ within a limited token budget, we select a representative subset $\mathcal{S}_j \subset C_j$. We adopt the same greedy sampling strategy as PersonaX, which balances \textit{prototypicality} and \textit{diversity} within each cluster. Prototypicality favors events closer to the cluster centroid, while diversity encourages coverage of heterogeneous behaviors.
\newline All sampling hyperparameters are set to the values reported in PersonaX, with a fixed sampling ratio of $0.6$ and a trade-off weight $\alpha = 1.06$.
Let $\mu_j$ be the centroid of cluster $C_j$. The scoring function for a candidate subset $\mathcal{S}_j$ is defined as:
\begin{equation}
    \mathcal{J}(\mathcal{S}_j) = w_p \sum_{e \in \mathcal{S}_j} \frac{1}{1 + \|\mathbf{v}_e - \mu_j\|} + w_d \cdot \frac{2}{|\mathcal{S}_j|} \sum_{e_a, e_b \in \mathcal{S}_j} \|\mathbf{v}_a - \mathbf{v}_b\|
\end{equation}
where $w_p = \alpha^{-10}$ and $w_d = 1 - w_p$. This mechanism ensures the selected events capture the cluster's core intent while covering heterogeneous behavioral patterns. 

\paragraph{Persona Generation.}
For each cluster, we prompt a large language model to generate a single textual persona $p_j$ that summarizes the user’s behavioral patterns represented by the cluster. The prompt instructs the model to abstract specific actions into habitual preferences (e.g., \textit{``User frequently consults API docs while coding''}) without revealing sensitive information. Each persona is constrained to $100\sim120$ tokens. This yields a persona bank $\mathcal{P}_u = \{p_1, \ldots, p_m\}$ for user $u$.

\paragraph{Persona Retrieval.}
At inference time, given a test context $x = (a, w, u, h)$, we retrieve the most relevant behavioral priors. We compute the cosine similarity between the query embedding $\phi(x)$ and each persona in $\mathcal{P}_u$. The top-$k$ ($k=5$ ) personas are retrieved to form the context set $\mathcal{P}^*$.

\paragraph{Prompt Augmentation.}
The retrieved personas are serialized and injected into the system prompt as explicitly labeled priors:
\begin{equation}
    \tilde{x} = x_{\text{context}} \oplus [\texttt{PERSONA}: \mathcal{P}^*] \oplus x_{\text{task}}
\end{equation}
The prompt structure and decoding strategy remain consistent with other baselines to ensure fair comparison.

\paragraph{Complexity.}
Persona construction requires a single embedding pass over historical events followed by hierarchical clustering, resulting in $O(N^2)$ (where $N$ is the history length) worst-case time per user but with small $N$ in practice. Persona retrieval at inference time scales linearly with the number of personas and introduces negligible overhead. This method introduces no additional trainable parameters and operates entirely at inference time.

\section{VLM Prompts}
\label{sec:vlm_prompts}

We design a structured prompting framework to process multimodal user activity data for proactive assistance prediction. Our approach supports three inference strategies: zero-shot, chain-of-thought (CoT)~\cite{wei2022chain}, and self-consistency~\cite{wang2023selfconsistency}, each with tailored prompt templates.

\subsection{Event Detection Prompt Templates}

The system prompt establishes the model's role as a screen activity monitor:

\begin{quote}
\textit{``You are an intelligent assistant responsible for monitoring user screen activity. Based on the user's current screen state and recent behavior, determine whether the user needs help from an AI assistant.''}
\end{quote}

The user prompt follows a hierarchical structure with four components:

\textbf{Recent Activity Context.} A summary of user activities from the preceding 5-minute window, providing temporal context for behavioral pattern recognition.

\textbf{Current State.} Structured metadata including application name, window title, timestamp, and a brief screen summary (truncated to 200 characters for efficiency).

\textbf{Screen Content.} For vision-enabled models (e.g., Qwen2.5-VL~\cite{wang2024qwen2vl}), we pass the screenshot directly with the marker ``[See attached image]''. For text-only models (e.g., Llama-3.1-8B-Instruct~\cite{dubey2024llama}), we provide OCR-extracted text from the current screen.

\textbf{Task Instruction.} The query section varies by inference method. For zero-shot prompting, we request direct binary prediction with intention classification. For CoT, we decompose the task into a four-step reasoning process. For self-consistency, we use the zero-shot format with multiple sampling and majority voting.

\subsection{Sequence Analysis and Intention Classification}

We define 16 intention categories covering common assistance scenarios (e.g., knowledge Q\&A, code programming, content creation, information retrieval). The model selects from this predefined taxonomy when predicting user intention, enabling consistent evaluation across methods.

For CoT prompting~\cite{wei2022chain}, we explicitly structure the reasoning chain into four steps:
\begin{enumerate}
    \item Describe what the user is currently doing
    \item Analyze potential problems or needs the user may encounter
    \item Judge whether AI assistance is needed (yes/no)
    \item Classify the user's intention category from the predefined set
\end{enumerate}

This decomposition encourages the model to ground its prediction in observable screen evidence before committing to a classification, following the principle that intermediate reasoning steps improve complex task performance~\cite{wei2022chain}.

For self-consistency~\cite{wang2023selfconsistency}, we sample multiple reasoning paths using temperature-based decoding and aggregate predictions via majority voting. This approach leverages the intuition that complex reasoning tasks typically admit multiple valid reasoning paths leading to the correct answer.

\subsection{Output Validation Rules}

We enforce structured outputs through explicit format specifications in the prompt. The expected response format is:

\begin{verbatim}
1. Need help: Yes/No
2. Intention category: [category]
\end{verbatim}

Response parsing employs regex-based extraction to handle format variations:
\begin{itemize}
    \item \textbf{Primary pattern matching}: We first search for explicit format adherence using the pattern \texttt{Need help: (Yes|No)}.
    \item \textbf{Fallback heuristics}: For responses that deviate from the template, we scan the raw response for keywords indicating the prediction.
    \item \textbf{Method-specific parsing}: For CoT responses, we additionally extract predictions from Step 3 of the reasoning chain when the final summary is malformed.
\end{itemize}

When the model's response does not conform to the expected format, we apply cascading rules: first searching for explicit markers, then scanning for category keywords with preference for later mentions (as CoT responses typically place final answers at the end). This robust parsing strategy ensures reliable evaluation even when models produce verbose or partially-formatted outputs.

\section{Full Results}

\subsection{Base Result \& Memory-based Methods Result}

\begin{figure*}[ht]
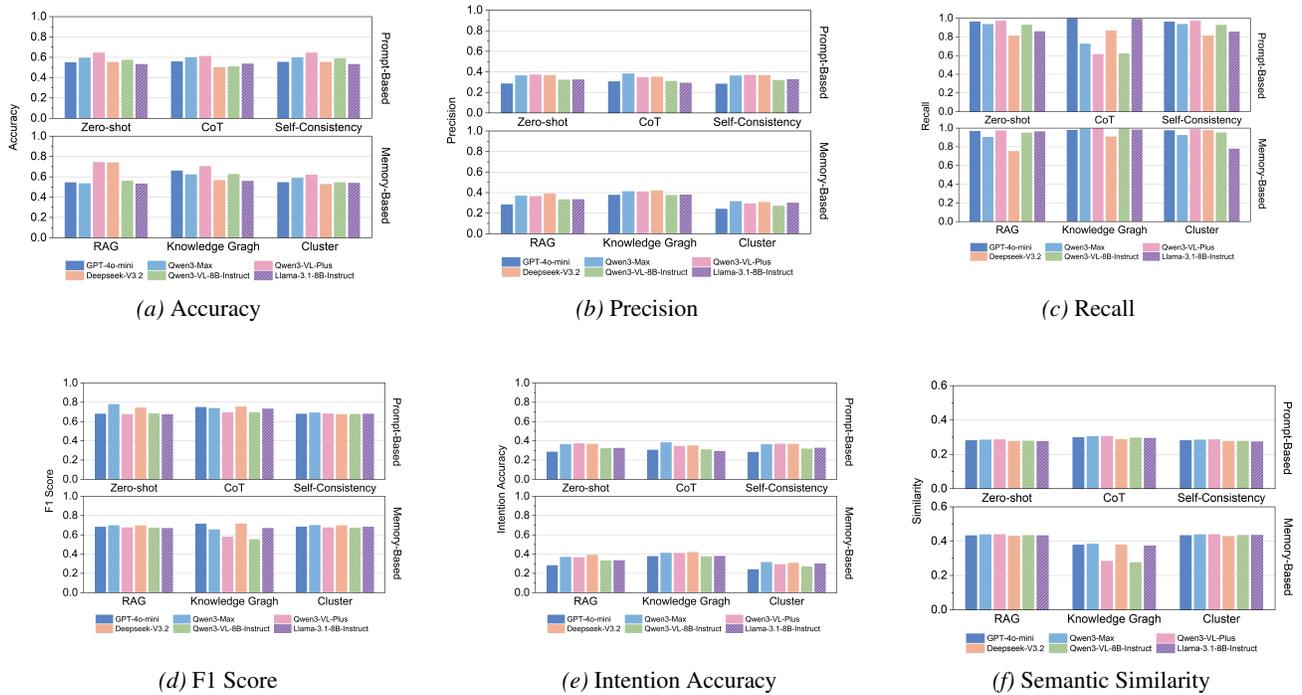

  \centering
  \begin{subfigure}[b]{0.3\textwidth}
    \includegraphics[width=\textwidth]{figures/Accuracy.jpg}
    \caption{Accuracy}
    \label{fig:base_acc}
  \end{subfigure}
  \hfill
  \begin{subfigure}[b]{0.30\textwidth}
    \includegraphics[width=\textwidth]{figures/Precision.jpg}
    \caption{Precision}
    \label{fig:base_precision}
  \end{subfigure}
  \hfill
  \begin{subfigure}[b]{0.32\textwidth}
    \includegraphics[width=\textwidth]{figures/recall.jpg}
    \caption{Recall}
    \label{fig:base_recall}
  \end{subfigure}
  
  \vspace{10pt}
  
  \begin{subfigure}[b]{0.33\textwidth}
    \includegraphics[width=\textwidth]{figures/F1_score.jpg}
    \caption{F1 Score}
    \label{fig:base_f1}
  \end{subfigure}
  \hfill
  \begin{subfigure}[b]{0.33\textwidth}
    \includegraphics[width=\textwidth]{figures/Intention_acc.jpg}
    \caption{Intention Accuracy}
    \label{fig:base_intention_acc}
  \end{subfigure}
  \hfill
  \begin{subfigure}[b]{0.3\textwidth}
    \includegraphics[width=\textwidth]{figures/simi.jpg}
    \caption{Semantic Similarity}
    \label{fig:base_simi}
  \end{subfigure}
  \caption{Base results and memory-based methods comparison across different evaluation metrics for both \textit{When to Assist} (Accuracy, Precision, Recall, F1 Score) and \textit{How to Assist} (Intention Accuracy, Semantic Similarity) tasks.}
  \label{fig:base_memory_results}
\end{figure*}

This section presents the complete evaluation results comparing base prompt-based methods (Zero-shot, CoT, Self-Consistency) with memory-augmented approaches (RAG, Knowledge Graph and Cluster). Figure~\ref{fig:base_memory_results} shows the performance across all six evaluation metrics: (a) Accuracy, (b) Precision, (c) Recall, and (d) F1 Score for the \textit{When to Assist} task, and (e) Intention Accuracy and (f) Semantic Similarity for the \textit{How to Assist} task. Each subplot compares different methods across multiple LLM backbones, demonstrating the effectiveness of incorporating long-term user context through memory mechanisms.

\subsection{Context Time Window Length Ablation}

\begin{figure*}[ht]
  \centering
  \begin{subfigure}[b]{0.32\textwidth}
    \includegraphics[width=\textwidth]{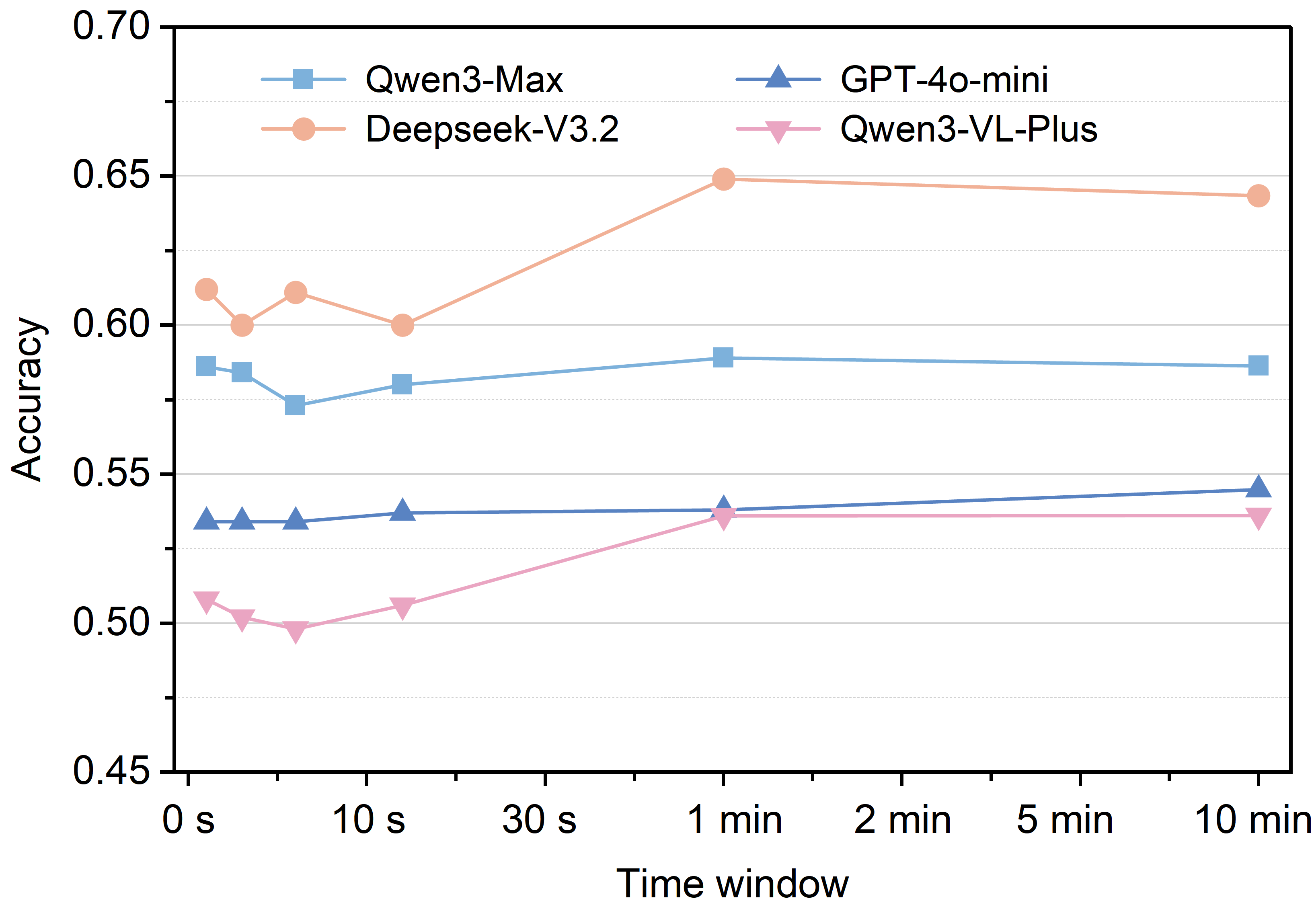}
    \caption{Accuracy}
    \label{fig:full_acc}
  \end{subfigure}
  \hfill
  \begin{subfigure}[b]{0.32\textwidth}
    \includegraphics[width=\textwidth]{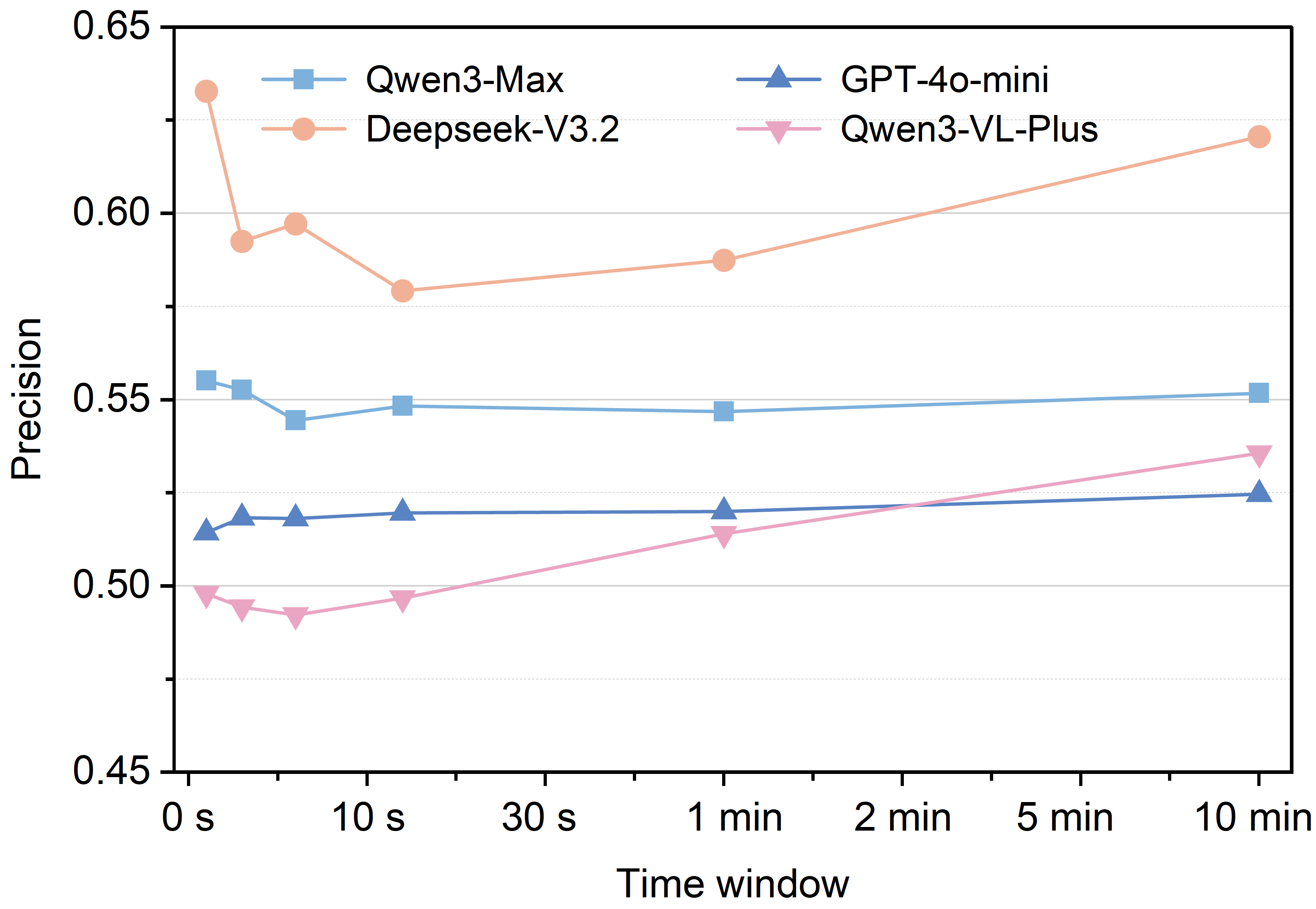}
    \caption{Precision}
    \label{fig:full_precision}
  \end{subfigure}
  \hfill
  \begin{subfigure}[b]{0.32\textwidth}
    \includegraphics[width=\textwidth]{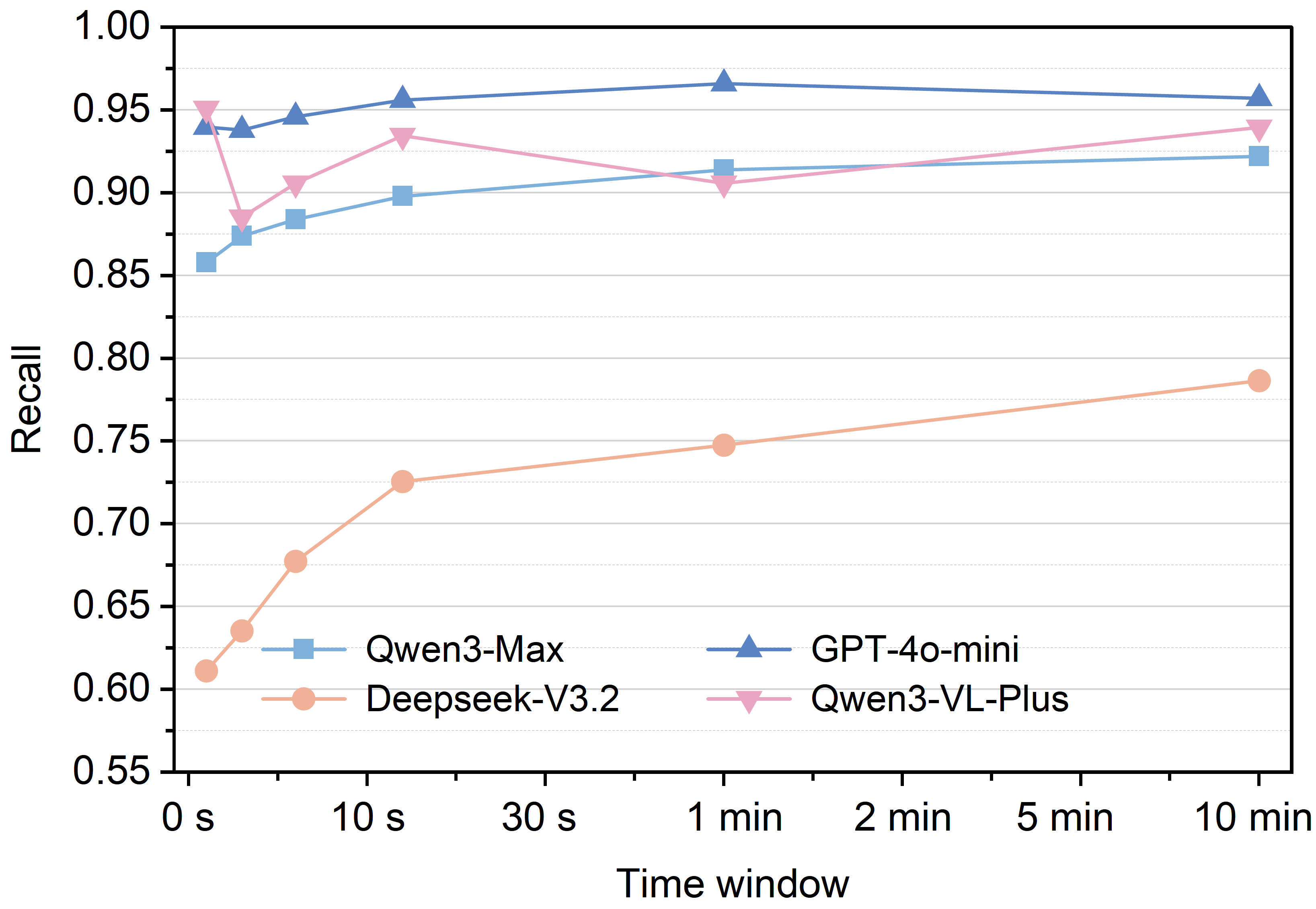}
    \caption{Recall}
    \label{fig:full_recall}
  \end{subfigure}
  
  \vspace{10pt}
  
  \begin{subfigure}[b]{0.32\textwidth}
    \includegraphics[width=\textwidth]{figures/time_window_F1.jpg}
    \caption{F1 Score}
    \label{fig:full_f1}
  \end{subfigure}
  \hfill
  \begin{subfigure}[b]{0.32\textwidth}
    \includegraphics[width=\textwidth]{figures/time_window_intention_acc.jpg}
    \caption{Intention Accuracy}
    \label{fig:full_intention_acc}
  \end{subfigure}
  \caption{Full evaluation results across different time window sizes (from 30s to 10m) for both \textit{When to Assist} and \textit{How to Assist} tasks.}
  \label{fig:full_results}
\end{figure*}

This section presents the ablation study on the impact of historical context length. We evaluate model performance across different time window sizes ranging from 30 seconds to 10 minutes. Figure~\ref{fig:full_results} shows the results: (a) Accuracy, (b) Precision, (c) Recall, and (d) F1 Score for the \textit{When to Assist} task, and (e) Intention Accuracy for the \textit{How to Assist} task. The results demonstrate that longer context windows generally improve performance, with diminishing returns observed beyond the 5-minute mark, suggesting that a 5-minute context window provides an effective balance between capturing sufficient behavioral context and computational efficiency.

\subsection{Real-world vs. Synthetic Training Data}

This section presents the complete results comparing models fine-tuned on real-world data versus synthetic data. Table~\ref{tab:real_world_data_full} shows the performance of LLaMA-3.1-8B-Instruct and Qwen3-VL-8B-Instruct under different fine-tuning strategies (SFT and LoRA) with both data sources. The results demonstrate that real-world data consistently outperforms synthetic data across all metrics, highlighting the unique value of authentic human interaction patterns for training proactive assistance systems.

\begin{table*}[ht]
  \centering
  \caption{Impact of training data source on open-source models (Full Results). We compare Zero-shot baseline with models fine-tuned on real-world vs. synthetic data using SFT and LoRA. Abbreviations: Acc.=Accuracy, Pre.=Precision, Rec.=Recall, Int. Acc.=Intention Accuracy, Sem. Sim.=Semantic Similarity. Best results per model are \textbf{bolded}.}
  \label{tab:real_world_data_full}
  \begin{tabular}{llccccccc}
    \toprule
    \textbf{Method} & \textbf{Data} & \textbf{Accuracy} & \textbf{Precision} & \textbf{Recall} & \textbf{F1 Score} & \textbf{Intention Accuracy} & \textbf{Semantic Similarity} \\
    \midrule
    \rowcolor{gray!15} \multicolumn{8}{l}{\textbf{LLaMA-3.1-8B-Instruct}} \\
    Zero-shot & -- & 57.3\% & 54.7\% & 85.7\% & 66.7\% & 32.3\% & 0.275 \\
    \cmidrule(lr){1-8}
    \multirow{2}{*}{SFT} & Synthetic & 62.1\% & 58.3\% & 86.2\% & 70.2\% & 34.8\% & 0.312 \\
                         & Real-world & \textbf{74.0\%} & \textbf{71.2\%} & \textbf{87.4\%} & \textbf{78.5\%} & \textbf{42.1\%} & \textbf{0.385} \\
    \cmidrule(lr){1-8}
    \multirow{2}{*}{LoRA} & Synthetic & 60.5\% & 57.1\% & 85.8\% & 68.9\% & 33.6\% & 0.298 \\
                          & Real-world & 71.2\% & 68.5\% & 86.1\% & 76.3\% & 40.5\% & 0.372 \\
    \midrule
    \rowcolor{gray!15} \multicolumn{8}{l}{\textbf{Qwen3-VL-8B-Instruct}} \\
    Zero-shot & -- & 51.7\% & 50.9\% & 94.4\% & 66.1\% & 35.3\% & 0.276 \\
    \cmidrule(lr){1-8}
    \multirow{2}{*}{SFT} & Synthetic & 54.8\% & 52.6\% & 92.8\% & 67.8\% & 36.2\% & 0.295 \\
                         & Real-world & \textbf{63.5\%} & \textbf{60.2\%} & \textbf{89.5\%} & \textbf{72.4\%} & \textbf{41.8\%} & \textbf{0.358} \\
    \cmidrule(lr){1-8}
    \multirow{2}{*}{LoRA} & Synthetic & 53.2\% & 51.8\% & 93.2\% & 67.1\% & 35.9\% & 0.288 \\
                          & Real-world & 61.8\% & 58.9\% & 90.1\% & 71.2\% & 40.2\% & 0.345 \\
    \bottomrule
  \end{tabular}
\end{table*}

\section{Ablations}

\subsection{Ablations 1: Impact of Agent Reasoning Strategies}
\label{sec:ablation_reasoning}
To explore the potential of advanced reasoning in proactive assistance, we evaluate different strategies including: (1) \textbf{Zero-shot}, the baseline approach that generates proactive decisions directly from input user information without explicit reasoning processes; (2) \textbf{Chain-of-Thought (CoT)}, which elicits step-by-step reasoning; and (3) \textbf{Self-Consistency (SC)}, which aggregates multiple inference paths.

We observe that: (1) SC is the most reliable prompting method, offering consistent performance compared to the baseline. For example, as shown in Table~\ref{tab:multimodal_performance_comparison}, in Qwen3-VL-8B-Instruct (Text-only), while CoT suffers a drastic performance drop (F1 Score: 22.4\%), SC maintains robust performance (F1 Score: 66.7\%), closely matching and slightly outperforming the Zero-shot baseline (F1 Score: 66.1\%), effectively mitigating the volatility seen in complex reasoning chains; (2) Interestingly, CoT prompting often yields lower performance than zero-shot. This aligns with recent findings that CoT degrades performance on tasks involving implicit pattern recognition rather than explicit logical deduction \cite{liu2025mind, zheng2025curse}. Our analysis reveals that CoT amplifies models' inherent behavioral tendencies: in Deepseek-V3.2 and LLaMA3.1-8B, CoT shifts decision boundaries toward aggressive triggering (higher Recall, lower Accuracy), while in Qwen3-VL-8B, it induces excessive conservatism (Recall drops from 0.944 to 0.171).
We further observe that CoT tends to overthink simple scenarios, imagining future problems rather than assessing what the user actually needs in the present. As illustrated in Figure~\ref{fig:cot_failure}, given a user simply browsing multiple tabs, zero-shot correctly predicts no assistance is needed. Under CoT, however, the model constructs unfounded reasoning about information overload and hypothetical needs to compare page contents, ultimately producing an incorrect prediction. Consequently, for proactive assistance systems where balancing false alarms and coverage is critical, zero-shot or Self-Consistency prompting remains the more robust choice.

\subsection{Ablations 2: Impact of Input Modalities}

\begin{table*}[t]
  \centering
  \caption{Performance of Multi-modal Models on When to Offer Assistance (Text-only vs. Multi-modal Inputs)}
  \label{tab:multimodal_performance_comparison}
  \begin{tabular}{llcccccc}
    \toprule
    \toprule
    \multirow{2}{*}{\textbf{Input}} & \multirow{2}{*}{\textbf{Metric}} & \multicolumn{3}{c}{\textbf{Prompt-based Methods}} & \multicolumn{3}{c}{\textbf{Memory-based Methods}} \\
     \cmidrule(lr{2pt}){3-5} \cmidrule(lr{2pt}){6-8}
     & & Zero-shot & CoT & SC & RAG & KG & Cluster \\
    \midrule
    \midrule
    \multicolumn{8}{c}{\textbf{Close-source Models}}\\
    \midrule
    \rowcolor{gray!15} \multicolumn{8}{l}{\textbf{GPT-4o-mini}} \\
    \multirow{6}{*}{\textit{Text-only}} 
      & Accuracy  & 54.9\% & 55.7\% & 55.2\% & 54.3\% & 65.9\% & 54.4\% \\
    & Precision & 52.7\% & 55.6\% & 52.8\% & 52.3\% & 60.5\% & 52.0\% \\
    & Recall    & 96.2\% & 99.5\% & 96.0\% & 96.6\% & 97.7\% & 97.4\% \\
    & F1 Score  & 68.1\% & 71.3\% & 68.2\% & 67.9\% & 74.8\% & 67.8\% \\
    & Intention Acc. & 28.4\% & 30.5\% & 28.2\% & 28.2\% & 37.6\% & 24.0\% \\
    & Semantic Sim. & 0.280 & 0.298 & 0.280 & 0.432 & 0.378 & 0.433 \\
    \cmidrule{1-8}
    \multirow{6}{*}{\textit{Multi-modal}}
      & Accuracy  & 52.5\% & 49.8\% & 53.0\% & 67.3\% & 62.2\% & 54.0\% \\
    & Precision & 51.3\% & 49.7\% & 51.6\% & 60.7\% & 57.0\% & 52.0\% \\
    & Recall    & 96.0\% & 99.8\% & 95.2\% & 97.1\% & 98.0\% & 97.3\% \\
    & F1 Score  & 66.9\% & 66.4\% & 66.9\% & 74.7\% & 72.1\% & 67.8\% \\
    & Intention Acc. & 34.0\% & 33.4\% & 33.3\% & 29.3\% & 40.4\% & 23.1\% \\
    & Semantic Sim. & 0.282 & 0.299 & 0.279 & 0.433 & 0.376 & 0.432 \\
    \midrule
    \rowcolor{gray!15} \multicolumn{8}{l}{\textbf{Qwen3-VL-Plus}} \\
    \multirow{6}{*}{\textit{Text-only}}
      & Accuracy  & 53.0\% & 53.5\% & 53.1\% & 53.3\% & 55.7\% & 53.7\% \\
    & Precision & 51.6\% & 54.9\% & 51.7\% & 51.8\% & 53.0\% & 51.9\% \\
    & Recall    & 97.0\% & 61.3\% & 97.0\% & 97.1\% & 99.4\% & 98.6\% \\
    & F1 Score  & 67.4\% & 57.9\% & 67.4\% & 67.5\% & 69.2\% & 68.0\% \\
    & Intention Acc. & 37.1\% & 34.4\% & 36.7\% & 36.3\% & 40.9\% & 29.2\% \\
    & Semantic Sim. & 0.286 & 0.305 & 0.286 & 0.439 & 0.284 & 0.439 \\
    \cmidrule{1-8}
    \multirow{6}{*}{\textit{Multi-modal}}
      & Accuracy  & 50.6\% & 46.7\% & 50.5\% & 55.7\% & 63.9\% & 53.4\% \\
    & Precision & 50.3\% & 45.8\% & 50.3\% & 53.2\% & 62.5\% & 51.9\% \\
    & Recall    & 94.2\% & 35.9\% & 93.4\% & 95.0\% & 99.7\% & 97.3\% \\
    & F1 Score  & 65.6\% & 40.3\% & 65.4\% & 68.2\% & 76.8\% & 67.7\% \\
    & Intention Acc. & 38.2\% & 38.0\% & 38.5\% & 38.1\% & 44.2\% & 31.0\% \\
    & Semantic Sim. & 0.286 & 0.296 & 0.285 & 0.439 & 0.384 & 0.438 \\
    \midrule
    \midrule
    \multicolumn{8}{c}{\textbf{Open-source Models}}\\
    \midrule
    \rowcolor{gray!15} \multicolumn{8}{l}{\textbf{Qwen3-VL-8B-Instruct}} \\
    \multirow{6}{*}{\textit{Text-only}}
      & Accuracy  & 51.7\% & 41.0\% & 52.9\% & 55.9\% & 56.5\% & 52.6\% \\
    & Precision & 50.9\% & 32.7\% & 51.8\% & 53.3\% & 53.5\% & 51.4\% \\
    & Recall    & 94.4\% & 17.1\% & 93.6\% & 94.9\% & 99.1\% & 97.6\% \\
    & F1 Score  & 66.1\% & 22.4\% & 66.7\% & 68.2\% & 69.5\% & 67.3\% \\
    & Intention Acc. & 35.3\% & 34.1\% & 35.7\% & 33.2\% & 37.5\% & 27.0\% \\
    & Semantic Sim. & 0.276 & 0.277 & 0.274 & 0.434 & 0.277 & 0.435 \\
    \cmidrule{1-8}
    \multirow{6}{*}{\textit{Multi-modal}}
      & Accuracy  & 51.7\% & 41.0\% & 52.9\% & 65.9\% & 54.2\% & 52.2\% \\
    & Precision & 50.9\% & 32.7\% & 51.8\% & 60.3\% & 52.2\% & 51.2\% \\
    & Recall    & 94.4\% & 17.1\% & 93.6\% & 93.7\% & 99.0\% & 98.2\% \\
    & F1 Score  & 66.1\% & 22.4\% & 66.7\% & 73.4\% & 68.4\% & 67.3\% \\
    & Intention Acc. & 35.3\% & 34.1\% & 35.7\% & 32.5\% & 41.6\% & 28.0\% \\
    & Semantic Sim. & 0.276 & 0.295 & 0.274 & 0.434 & 0.376 & 0.436 \\
    \bottomrule
    \bottomrule
  \end{tabular}
\end{table*}

To investigate whether visual context improves proactive assistance, we evaluate model performance across two input modalities: \textbf{Multi-modal} and \textbf{Text-only}. In the Multi-modal setting, the model receives the raw screen screenshot, combined with the user's historical interaction data and profile information. In contrast, the Text-only setting replaces the visual screenshot with its textual representation, extracted via Optical Character Recognition (OCR), while retaining the same user history and profile context.

We observe that: (1) Surprisingly, integrating visual information does not consistently improve performance and, in many cases, leads to degradation. For instance, in Qwen3-VL-Plus (Table~\ref{tab:multimodal_performance_comparison}), the Multi-modal input yields lower Accuracy (50.6\% vs. 53.0\%), F1 Score (65.6\% vs. 67.4\%), and Precision (50.3\% vs. 51.6\%) compared to the Text-only baseline in the Zero-shot setting. A similar trend is observed in GPT-4o-mini, where Multi-modal accuracy drops to 52.5\% from 54.9\% (Text-only); (2) Text-only models demonstrate greater stability and efficiency. Across most models and prompting strategies (e.g., Qwen3-VL-8B-Instruct with SC), Text-only inputs achieve comparable or identical F1 Scores (66.7\%) to their Multi-modal counterparts, suggesting that current VLMs may struggle to effectively extract actionable proactive cues from complex GUI screenshots, or that the essential context is already sufficiently captured by the text logs.

\subsection{Ablations 3: Inference Latency of Different Methods}

\begin{figure}[t]
  \centering
  \includegraphics[width=0.8\linewidth]{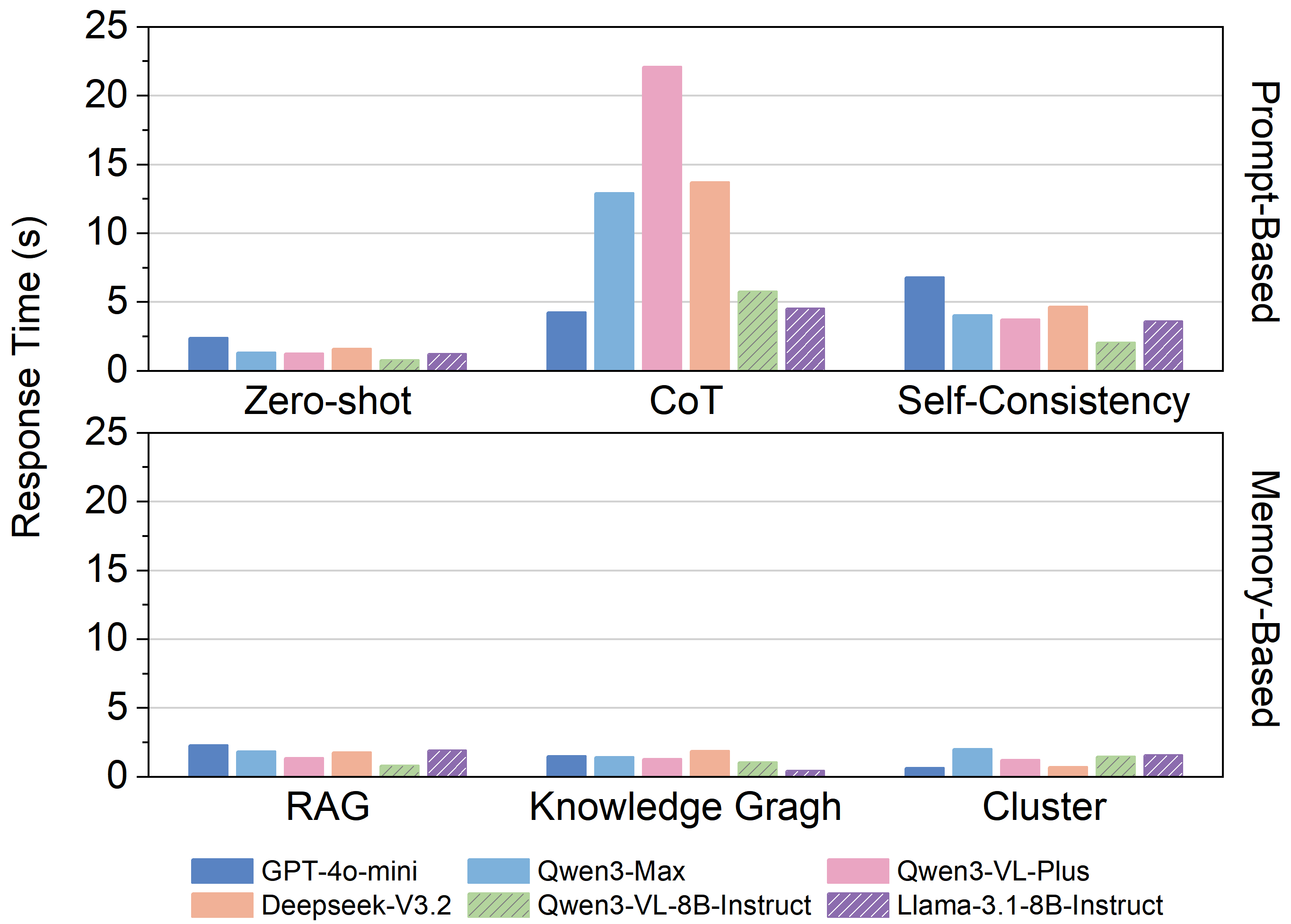}
  \caption{\textbf{Inference latency comparison across methods.} Response time (in seconds) for different models using prompt-based methods (top) and memory-based methods (bottom).}
  \label{fig:inference_time}
\end{figure}

For real-world deployment of proactive assistance systems, inference latency is a critical factor, as excessive delays can diminish user experience and reduce the practical utility of timely interventions. We measure the average response time across all evaluated methods and models, as shown in Figure~\ref{fig:inference_time}.

We observe that: (1) Most methods achieve real-time or near-real-time inference. Zero-shot prompting demonstrates the lowest latency, with most models responding within 5 seconds, making it highly suitable for latency-sensitive applications. Memory-based methods (RAG, Knowledge Graph, Clustering) also maintain low latency ($<$2 seconds), as the retrieval and reasoning overhead is minimal compared to generation; (2) \textbf{Chain-of-Thought (CoT) substantially increases inference time.} Across all models, CoT introduces significant latency overhead due to the explicit multi-step reasoning process. For instance, Qwen3-VL-Plus requires approximately 22 seconds, while Deepseek-V3.2 and Qwen3-Max take around 13-14 seconds, which is an order of magnitude slower than Zero-shot. This latency penalty, combined with CoT's inconsistent performance improvements (Section~\ref{sec:ablation_reasoning}), suggests that CoT may not be the optimal choice for proactive assistance scenarios where rapid response is essential; (3) Self-Consistency exhibits moderate latency (3-7 seconds), as it requires multiple sampling passes. Given its stable performance and acceptable latency trade-off, Self-Consistency represents a reasonable middle ground for applications that can tolerate slightly longer response times.

\section{CoT Failure Cases}

We observe that CoT prompting yields mixed results depending on model capacity. For larger models such as Deepseek-V3.2, CoT improves the F1 score from 69.5\% to 71.3\% on timing prediction. However, for smaller open-source models, CoT can be detrimental. For example, Qwen3-VL-8B-Instruct experiences a dramatic performance drop, with accuracy falling from 51.7\% to 41.0\%. This aligns with recent findings that CoT degrades performance on tasks involving implicit pattern recognition rather than explicit logical deduction \cite{liu2025mind, zheng2025curse}. Our analysis reveals that CoT amplifies models' inherent behavioral tendencies: in Deepseek-V3.2 and LLaMA3.1-8B, CoT shifts decision boundaries toward aggressive triggering (higher Recall, lower Accuracy), while in Qwen3-VL-8B, it induces excessive conservatism (Recall drops from 0.944 to 0.171).
We further observe that CoT tends to overthink simple scenarios, imagining future problems rather than assessing what the user actually needs in the present,  as illustrated in Figure~\ref{fig:cot_failure}.
On the \textit{How to Assist} task, CoT provides modest improvements in semantic similarity (e.g., Qwen3-Max improves from 0.285 to 0.305), indicating that structured reasoning helps models better articulate assistance content.

As is demonstrated in Fig. \ref{fig:cot_failure}, given a user simply browsing multiple tabs, zero-shot correctly predicts no assistance is needed. Under CoT, however, the model constructs unfounded reasoning about information overload'' and hypothetical needs to compare page contents,'' ultimately producing an incorrect prediction. Consequently, for proactive assistance systems where balancing false alarms and coverage is critical, zero-shot or Self-Consistency prompting remains the more robust choice.
\begin{figure}[t]
  \centering
  \footnotesize
  \setlength{\tabcolsep}{4pt}
  
  \begin{tabular}{@{}p{0.08\linewidth}p{0.62\linewidth}p{0.22\linewidth}@{}}
  \toprule
  \multicolumn{3}{l}{\textbf{Zero-shot Prediction}} \\
  \midrule
  \textit{Output} &
  ``Need help: \textbf{No}; Intent category: \textbf{Uncategorized}.'' &
  \textcolor{green!60!black}{\textbf{\ding{51}{ Correct}}} \\
  \bottomrule
  \end{tabular}
  
  \vspace{3mm}
  
  \begin{tabular}{@{}p{0.08\linewidth}p{0.62\linewidth}p{0.22\linewidth}@{}}
  \toprule
  \multicolumn{3}{l}{\textbf{Chain-of-Thought Prediction}} \\
  \midrule
  \textbf{Step} & \textbf{Reasoning Path} & \textbf{Assessment} \\
  \midrule
  S1 &
  ``The user is currently using the Microsoft Edge browser with multiple tabs open, including a page related to Google Gemini. The user has been active in this browser during the past five minutes.'' &
  \textcolor{green!60!black}{Observed context} \\
  \midrule
  S2 &
  ``The user may be browsing information about Google Gemini or conducting online activities such as research or study. However, since no explicit text content is available, the exact activity cannot be determined.'' &
  \textcolor{orange}{Speculation with uncertainty} \\
  \midrule
  S3 &
  ``Because the user has multiple pages open, they may experience \underline{information overload}, need to \underline{quickly locate specific information}, or \underline{compare content across tabs}. Therefore, the user may require assistance.'' &
  \textcolor{red}{Unfounded inference} \\
  \midrule
  S4 &
  ``The user's intent is likely \emph{Information Query}, as they may be searching for Gemini-related information or comparing multiple pages.'' &
  \textcolor{red}{Intent inferred from fabrication} \\
  \midrule
  Final &
  ``Need help: \textbf{Yes}; Intent: \textbf{Information Query}.'' &
  \textcolor{red}{\textbf{\ding{55}{ Incorrect}}} \\
  \bottomrule
  \end{tabular}
  
  \caption{Illustrative false-positive case of Chain-of-Thought (CoT) reasoning in proactive help prediction.
  \textbf{Context:} The user is browsing multiple tabs in Microsoft Edge.
  \textbf{Ground truth:} No assistance is required.
  While the zero-shot model directly outputs the correct decision, the CoT model progressively introduces hypothetical user difficulties (underlined in Step~S3), leading to an incorrect prediction of help need and intent.}
  \label{fig:cot_failure}
  \end{figure}

\end{document}